\documentclass[12pt,preprint]{aastex}
\usepackage{color}
\shorttitle{Spectral-Temporal Simulations of GRBs}
\shortauthors{Asano \& M\'esz\'aros}

\begin{document}

\title{
Spectral--Temporal Simulations of 
Internal Dissipation Models of Gamma-Ray Bursts
}
\author{\scshape Katsuaki Asano\altaffilmark{1},
and
Peter M\'esz\'aros\altaffilmark{2}}
\email{asano@phys.titech.ac.jp, nnp@astro.psu.edu}

\altaffiltext{1}{Interactive Research Center of Science, %Graduate School of Science,
Tokyo Institute of Technology, 2-12-1 Ookayama, Meguro-ku, Tokyo 152-8550, Japan}
\altaffiltext{2}{Department of Astronomy \& Astrophysics;
Department of Physics;
Center for Particle Astrophysics;
Pennsylvania State University,
University Park, PA 16802}

\date{Submitted; accepted}

\begin{abstract}
We present calculations of the time evolution of the prompt spectra of
gamma-ray burst models
involving generic internal dissipation regions, including internal shocks, either
by itself or in the presence of an external photon source such as a photosphere.
The method uses a newly developed time-dependent code involving 
synchrotron emission and absorption, inverse Compton scattering and pair formation. 
The models reproduce the typically observed Band spectra and their generic
time evolution, including the appearance of an extra keV-GeV component, whose
delay in simple SSC models, however, is only partially able to explain the
several seconds observed GeV delays. On the other hand, models involving both a 
photosphere and an internal dissipation region at a larger radius produce both
an extra GeV component and time delays which are in the range of the observations.

\end{abstract}


\maketitle

\section{Introduction}
\label{sec:intro}

Most of the prompt emission spectra of gamma-ray bursts (GRBs)
can be described by the well-known Band function \citep{ban93},
which has a spectral peak around the MeV range.
Since the EGRET era \citep{gon03}, distinct spectral components
in higher energy ranges have been reported.
Moreover, bright optical emissions during the prompt phase,
such as in GRB 080319B \citep{rac08}, have also been detected.
Such lower energy emissions also consist of distinct spectral components.
In the popular internal shock model of GRB the prompt MeV gamma-rays 
are attributed to synchrotron radiation from electrons accelerated in shocks 
within the GRB outflow \citep{pir05,mes06}, while higher energy photons are 
attributed to inverse Compton (IC) upscattering of these synchrotron photons, a 
scheme called synchrotron self-Compton or SSC \citep{mes94,pil98,pan00,gue03}.
Variants on this scheme involve having the MeV photons, whether of synchrotron or 
other origin, arising at a different location (usually at a smaller radius) than 
the location where they are upscattered, a scheme refereed to as external inverse 
Compton or EIC \citep{bel05,tom09,mur11}.
The generic radiation physics is independent, 
to a large degree, of the specific macroscopic model for the acceleration of 
the radiating or scattering non-thermal particles. Thus, the results we discuss 
here are not tied specifically to internal shock models, but apply also to more 
generic dissipative models, in which synchrotron, IC and, in the case
of high radiation density also pair formation, are the main radiation mechanisms.

The recent observations over a wide range of photon energies
by the {\it Fermi}-Large Area Telescope (LAT) and GLAST Burst Monitor
(GBM) instruments and
robotic optical telescopes such as TORTORA have stimulated research on
the temporal evolution of the GRB spectra,
which is a key for distinguishing various models.
Several authors have developed time-dependent codes and discuss 
the spectral evolution in GRBs \citep{pee05,pee08,bel08,vur09,bos09,dai11}.
Some of those studies focus on the effects of the dynamical evolution
of the shocked region, multiple Compton scattering, or thermal particles etc.
An expanding range of physical situations needs to be considered, in
view of the ongoing observational and theoretical progress being made.
Here, we improve our previous Monte Carlo code, developed in a series of
studies for hadronic and leptonic processes \citep[see, e.g.,][]{asa07},
into a time-dependent code. Our code is based on the one-zone approximation
with the Lagrangian specification in the energy space for particles.
Although we plan to develop this code to deal with the hadronic processes in 
the future, at this time only the leptonic processes are included, as a first step.

Our goal in this paper is to study the temporal behavior of the radiation field 
in the typical types of geometries generally considered in GRB models. 
In particular, we study the development of
the photon spectra as a function of time, as well as the expected light curves
in specific energy ranges, in order to gain insight into  current questions
such as the origin of the delay between MeV and GeV pulses in bursts observed
by the {\it Fermi} \citep{916C,902B,926A},
or the presence of additional
spectral components above or below \citep{510,902B,926A}
the usual Band (MeV) spectra known 
from previous Compton and {\it Swift} studies.

In \S \ref{sec:model} we present the basic physical model used in our
numerical, discussing the various physical processes involved and the way the
time-dependent spectra and light curves are computed. The results are presented
in \S \ref{sec:results}. In subsection \S \ref{sec:moderate} we discuss first 
three cases characterized by parameters such as inferred from pre-Fermi GRB 
observations, in the context of an SSC model. These ``moderate'' models serve 
as a reference point, providing insight into what was missed in the pre-Fermi
observations.  In subsection \S \ref{sec:lat} we explore the temporal and spectral 
properties and predictions for bursts with parameters which might characterize
some of the more interesting bursts observed by the Fermi LAT instrument, again 
based on an SSC model. In \S \ref{sec:external} we discuss the properties of a 
model including external photon sources providing a Band function spectrum which 
is reprocessed and upscattered at a different location.  We summarize our numerical 
results in \S \ref{sec:sum} and discuss their relevance and implications in 
\S \ref{sec:disc}.

\section{Model and Methods}
\label{sec:model}

In order to simulate the temporal evolution of photon emissions,
we develop a new numerical code by improving our code matured
via a series of GRB studies \citep{asa05,asa06,asa07,asa09},
in which the photon spectrum had been obtained with the steady-state approximation
for the photon field.

Here we consider a shell expanding with the Lorentz factor $\Gamma$
from an initial radius $R=R_0$. The calculation of the photon production
is carried out in the shell frame (hereafter, the quantities in this frame
are denoted with primed characters). 
This shell represents the internal dissipation region, which in what 
follows will be referred to as ``internal shock", although it could be a 
generic dissipation region resulting in particle acceleration, e.g., a
magnetic reconnection zone, etc.
Let us adopt notations for particle kinetic energies of electrons and photons
as $\varepsilon_{\rm e}=(\gamma_{\rm e}-1) m_{\rm e} c^2$ and
$\varepsilon_\gamma$, respectively.
The one-zone approximation with a constant shell width
$W'=R_0/\Gamma$ is adopted so that the photon density
$n'_\gamma(\varepsilon'_\gamma)$ and magnetic field
$B'$, etc., in the shell are homogeneous.
We refresh the photon field with a time step $\Delta T'=R/(100 \Gamma c)$,
during which electrons/positrons emit photons in the constant photon field.
With the same time step we inject electrons intermittently.
In this paper, we assume a constant injection rate with total energy
$E_{\rm e}=\Gamma E'_{\rm e}$ during a timescale $t'_{\rm inj}=W'/c$.
The injection spectrum is assumed as a cut-off power-law shape,
$\dot{N'}_{\rm e,inj}(\varepsilon'_{\rm e})
\propto \varepsilon'^{-p}_{\rm e} \exp{(-\varepsilon'_{\rm e}/\varepsilon'_{\rm max})}$
for $\varepsilon'_{\rm e} > \varepsilon'_{\rm min}$, where
$\varepsilon'_{\rm max}$ is determined by equating the cooling timescale $t'_{\rm cool}$
and the acceleration timescale
$t'_{\rm acc}=\xi \varepsilon'_{\rm e}/c e B'$
(hereafter, the efficient acceleration of $\xi=1$ is supposed).
Using the cooling rates due to synchrotron ($\dot{\cal E}_B$) and
IC ($\dot{\cal E}_{\rm IC}$), and the heating
rate due to synchrotron self absorption (SSA) ($\dot{\cal E}_{\rm SA}$),
the cooling time is defined as $t_{\rm cool}=\varepsilon_{\rm e}/
(\dot{\cal E}_B+\dot{\cal E}_{\rm IC}-\dot{\cal E}_{\rm SA})$.

The cooling timescale would be much shorter than the time step $\Delta T'$.
An even shorter time step $\Delta t'=\min(\Delta T'/10,t'_{\rm cool}/100)$
is used to evolve the electron energy distribution during $\Delta T'$
by following each electron with the Lagrangian specification in the energy space.
In every time step $\Delta t'$, we accumulate emitted photons,
and add them to the photon field $n'_\gamma(\varepsilon'_\gamma)$
in every $\Delta T'$.
In the following subsection we explain the detailed methods to
estimate $t'_{\rm cool}$ and simulate the emission processes.
The methods are essentially the same as our old code.

\subsection{Physical processes}
\label{sec:procs}

\subsubsection{Synchrotron emission}
\label{sec:syn}

The energy loss rate due to synchrotron/cyclotron emission is
%%%%%%%%%%%%%%%%%%%%%%%
\begin{eqnarray}
\dot{\cal E}_B=
\frac{1}{4 \pi}\sigma_{\rm T} c B^2 \gamma^2_{\rm e} \beta^2_{\rm e} \sin^2{\alpha},
\end{eqnarray}
%%%%%%%%%%%%%%%%%%%%%%%
where $\sigma_{\rm T}$ is the Thomson cross section,
and $\alpha$ is the pitch angle \citep{ryb79}.
In the ultra-relativistic limit ($\gamma_{\rm e} \gg 1$),
the energy spectrum of synchrotron emission
per unit time per photon energy is written as
%%%%%%%%%%%%%%%%%%%%%%%
\begin{eqnarray}
P_B(\varepsilon_{\rm e},\varepsilon_\gamma)&=&
\frac{\sqrt{3} e^3 B \sin{\alpha}}{2 \pi \hbar m_{\rm e} c^2} F(x), \\
x &\equiv& \frac{2 m_{\rm e} c \varepsilon_\gamma}{3 \gamma^2_{\rm e}
\hbar e B \sin{\alpha}},
\end{eqnarray}
%%%%%%%%%%%%%%%%%%%%%%%
where $F(x)$ is the synchrotron function.
In the non-relativistic energy range, instead of the exact formula
\citep[see, e.g.,][]{pee05},
we adopt a simple
approximation \citep{ghi98} for $\beta_{\rm e}=\sqrt{1-1/\gamma^2_{\rm e}}<0.89$ as
%%%%%%%%%%%%%%%%%%%%%%%
\begin{eqnarray}
P_B(\varepsilon_{\rm e},\varepsilon_\gamma)&=&
\frac{m_{\rm e} c^2 \sigma_{\rm T} B}{3 \pi \hbar e}
\frac{\gamma^2_{\rm e} \beta^2_{\rm e}}{1+3 \gamma^2_{\rm e} \beta^2_{\rm e}}
\exp{\left[ \frac{2 (1-\varepsilon_\gamma/\varepsilon_B)}
{1+3 \gamma^2_{\rm e} \beta^2_{\rm e}} \right]}, \\
\varepsilon_B &\equiv& \frac{\hbar e B}{m_{\rm e} c},
\end{eqnarray}
%%%%%%%%%%%%%%%%%%%%%%%
for the isotropic distribution of electrons.

\subsubsection{Inverse Compton}

In order to calculate emission due to
photon scattering with electrons, we prepared tables in advance.
Since we plan to develop this code to simulate optically thick plasmas
in the future, the tables cover a wide energy range from the Thomson regime
to the Klein-Nishina limit.
In the laboratory frame, the photon distribution is assumed to be isotropic.
Integrating over incident and scattering angles with the Monte Carlo method,
we obtained the tables of the emission spectrum as below.
Here, we denote the quantities in the electron rest frame
by characters with tilde.
Given the energies of an electron $\varepsilon_{\rm e}$ and
an incident photon $\varepsilon_{\gamma, \rm in}$
and a cosine of incident angle $\mu_{\rm in}$, the Lorentz transformation
yields the photon energy in the electron rest frame
$\tilde{\varepsilon}_{\gamma, \rm in}$.
The energy of the scattered photon
depends on the cosine of the scattering angle $\tilde{\mu}$ as
%%%%%%%%%%%%%%%%%%%%%%%
\begin{eqnarray}
\tilde{\varepsilon}_{\gamma, \rm out} =
\frac{\tilde{\varepsilon}_{\gamma, \rm in}}
{1+(\tilde{\varepsilon}_{\gamma, \rm in}/m_{\rm e} c^2)
(1-\tilde{\mu})}.
\end{eqnarray}
%%%%%%%%%%%%%%%%%%%%%%%
The scattering probability against $\tilde{\mu}$ is
%%%%%%%%%%%%%%%%%%%%%%%
\begin{eqnarray}
{\cal P}_{\tilde{\mu}} \propto
\left( \frac{\tilde{\varepsilon}_{\gamma, \rm out}}
{\tilde{\varepsilon}_{\gamma, \rm in}} \right)^2
\left[ \frac{\tilde{\varepsilon}_{\gamma, \rm in}}
{\tilde{\varepsilon}_{\gamma, \rm out}}+
\frac{\tilde{\varepsilon}_{\gamma, \rm out}}
{\tilde{\varepsilon}_{\gamma, \rm in}}-
(1-\tilde{\mu}^2)
\right],
\end{eqnarray}
%%%%%%%%%%%%%%%%%%%%%%%
which should be normalized by the total cross section as
%%%%%%%%%%%%%%%%%%%%%%%
\begin{eqnarray}
\sigma_{\rm KN}&=&\frac{3}{4} \sigma_{\rm T}
\left[
\frac{1+x}{x^3} \left( \frac{2x(1+x)}{1+2x}-\ln(1+2x) \right)
+ \right. \nonumber
\\ && \left. \frac{\ln(1+2x)}{2x}-\frac{1+3x}{(1+2x)^2}
\right], \\
x &\equiv& \tilde{\varepsilon}_\gamma/m_{\rm e} c^2,
\end{eqnarray}
%%%%%%%%%%%%%%%%%%%%%%%
\citep{ryb79}.
Reverting to the laboratory frame again,
we obtain the tables of the emission spectrum normalized
by the photon density in a form of
%%%%%%%%%%%%%%%%%%%%%%%
\begin{eqnarray}
P_{\rm IC}(\varepsilon_{\rm e},\varepsilon_{\gamma, \rm in},
\varepsilon_{\gamma, \rm out})/n_\gamma(\varepsilon_{\gamma, \rm in}).
\end{eqnarray}
%%%%%%%%%%%%%%%%%%%%%%%
At the same time we obtain the tables of the ``photon-absorption'' probability
for incident photons,
which assures the conservation of the photon number in the scattering process.
The Monte Carlo integral over the angles
brings us another table of the average value,
%%%%%%%%%%%%%%%%%%%%%%%
\begin{eqnarray}
\overline{\Delta \varepsilon_\gamma \sigma_{\rm KN}} \equiv
\overline{(1-\beta_{\rm e} \mu_{\rm in})
(\varepsilon_{\gamma, \rm out}-\varepsilon_{\gamma, \rm in})\sigma_{\rm KN}},
\end{eqnarray}
%%%%%%%%%%%%%%%%%%%%%%%
which is useful to calculate the energy loss rate as
%%%%%%%%%%%%%%%%%%%%%%%
\begin{eqnarray}
\dot{\cal E}_{\rm IC}&=&c
\int d \varepsilon_\gamma 
n_\gamma(\varepsilon_\gamma)
\overline{\Delta \varepsilon_\gamma
\sigma_{\rm KN}(\varepsilon_{\rm e},\varepsilon_\gamma)} \nonumber \\
&=&\int d \varepsilon_\gamma d \varepsilon_{\gamma, \rm out}
P_{\rm IC}(\varepsilon_{\rm e},\varepsilon_\gamma,
\varepsilon_{\gamma, \rm out}).
\end{eqnarray}
%%%%%%%%%%%%%%%%%%%%%%%

\subsubsection{Synchrotron self-absorption}

Adopting the synchrotron/cyclotron formulae in \S \ref{sec:syn},
the cross section of SSA
\citep{ryb79,ghi91} is written as
%%%%%%%%%%%%%%%%%%%%%%%
\begin{eqnarray}
\sigma_{\rm SA}&=&\frac{c^2 h^3}{8 \pi \varepsilon_\gamma^3}
\left( \frac{\gamma^2_{\rm abs} \beta_{\rm abs}}{\gamma^2_{\rm e} \beta_{\rm e}}
P_B(\varepsilon_{\rm abs},\varepsilon_\gamma) -
P_B(\varepsilon_{\rm e},\varepsilon_\gamma) \right), \\
\varepsilon_{\rm abs}&=&(\gamma_{\rm abs}-1) m_{\rm e} c^2=
\varepsilon_{\rm e}+\varepsilon_\gamma,
\end{eqnarray}
%%%%%%%%%%%%%%%%%%%%%%%
A fraction $1-\exp{(-c \Delta t' \int d \varepsilon_{\rm e} n_{\rm e} \sigma_{\rm SA})}$
of photons will be absorbed every time step.
The accumulated amount of absorbed photons is counted in every $\Delta T'$.
The heating rate of electrons due to SSA is
%%%%%%%%%%%%%%%%%%%%%%%
\begin{eqnarray}
\dot{\cal E}_{\rm SA}=\int d \varepsilon_\gamma
c \varepsilon_\gamma \sigma_{\rm SA} n_\gamma(\varepsilon_\gamma).
\end{eqnarray}
%%%%%%%%%%%%%%%%%%%%%%%

\subsubsection{Electron-Positron pair creation}

In order to take into account $\gamma \gamma$-pair creation,
every time step $\Delta T'$ we eliminate a fraction of photons,
$1-\exp{(-\tau_\pm)}$, with
%%%%%%%%%%%%%%%%%%%%%%%
\begin{eqnarray}
\tau_\pm(\varepsilon_{\gamma,1})=
c \Delta T' \int d \varepsilon_{\rm e} d \Omega_{\rm in}
(1-\mu_{\rm in}) \frac{n_\gamma(\varepsilon_{\gamma,2})}{4 \pi} \sigma_{\pm},
\end{eqnarray}
%%%%%%%%%%%%%%%%%%%%%%%
where
%%%%%%%%%%%%%%%%%%%%%%%
\begin{eqnarray}
\sigma_{\pm}&=&\frac{3}{16} \sigma_{\rm T} (1-y^2)
\left[ (3-y^4) \ln{\frac{1+y}{1-y}}-2y (2-y^2)
\right], \\
y^2 &\equiv& 1-(2m_{\rm e}^2 c^4)/[\varepsilon_{\gamma,1} \varepsilon_{\gamma,2}
(1-\mu_{\rm in})]
\end{eqnarray}
%%%%%%%%%%%%%%%%%%%%%%%
\citep{ber82}.
The integral over the photon energy and incident solid angle $d \Omega_{\rm in}
=2 \pi d \mu_{\rm in}$
is numerically calculated with the isotropic photon distribution.
We simultaneously inject secondary electron/positron pairs.
To save computational cost, we adopt a simple approximation
for the energy of the secondary pairs as
$\varepsilon_{\rm e}=(\varepsilon_{\gamma,1}+\varepsilon_{\gamma,2})/2
-m_{\rm e} c^2$
and neglect the effect of electron-positron
pair annihilation.

\subsubsection{Shell expansion}

We take into account the effect of adiabatic cooling,
though it may be a minor effect in GRB physics.
The adiabatic cooling of the collisionless plasma
is not trivial. The motion of electrons/positrons is
controlled by only magnetic fields. If a homogeneous
magnetic field evolves according to volume expansion,
only the momentum component perpendicular to the magnetic field
would be affected.
However, the magnetic field in the GRB sources may be highly
entangled so that the pitch angle diffusion cannot be neglected.
Thus, in this paper, we use a simple isotropic approximation for
the ``momentum'',
$\sqrt{\varepsilon'^2_{\rm e}+2 m_{\rm e} c^2 \varepsilon'_{\rm e}} c^{-1}
\propto V'^{-1/3}$,
which is applicable from non-relativistic ($\varepsilon'_{\rm e} \propto V'^{-2/3}$)
to ultra-relativistic ($\varepsilon'_{\rm e} \propto V'^{-1/3}$) limits.
The constant shell width $W'$ assumed in this paper
implies that the volume evolves as $V'=4 \pi R^2 W' \propto R^2$.
The volume expansion may imply evolution of the magnetic field,
but we assume constant $B'$ for simplicity in this paper.

\subsubsection{Photon escape}

The photon density $n'_\gamma$ refreshed every time step
gives us the photon escape rate per unit surface as
$n'_\gamma c/4$. The photons escape from both the foreside and backside
surfaces of the shell (the total surface: $2 \times 4 \pi R^2$).
Every time step we extract a fraction of photons,
$2 \pi R^2 c \Delta T'/V'=c \Delta T'/(2 W')$.

\subsection{Light Curve}
\label{sec:LC}

We consider photons escaping from the shell
with latitudes $\theta \leq \theta_{\rm jet}$ to take into account
the jetted structure, but
the maximum latitude is taken as large as $\theta_{\rm jet} \equiv 5/\Gamma$
so that the effect of collimation does not affect observation very much.
The energy and escape time $t'$ of photons coming from a latitude $\theta$
are transformed into energy- and time-bins in observer's frame with
%%%%%%%%%%%%%%%%%%%%%%%
\begin{eqnarray}
\varepsilon_{\rm obs}=
\frac{\varepsilon'_\gamma}{\Gamma (1-\beta_{\rm sh} \cos{\theta}) (1+z)},
\label{E-LT}
\end{eqnarray}
%%%%%%%%%%%%%%%%%%%%%%%
and
%%%%%%%%%%%%%%%%%%%%%%%
\begin{eqnarray}
t_{\rm obs}=(1+z) \left[ (1-\beta_{\rm sh} \cos{\theta}) \Gamma t'
+R_0 (1-\cos{\theta})/c \right],
\label{T-LT}
\end{eqnarray}
%%%%%%%%%%%%%%%%%%%%%%%
respectively, where $\beta_{\rm sh} \equiv \sqrt{1-1/\Gamma^2}$.
Every time step $\Delta T'$ the spectral number of escaped photons
per unit solid angle and surface is estimated with the photon density $n'_\gamma$
as
%%%%%%%%%%%%%%%%%%%%%%%
\begin{eqnarray}
\frac{dN'(\varepsilon'_\gamma)}{d \Omega' dS'}=\left| \cos{\theta'} \right|
\frac{n'_\gamma(\varepsilon'_\gamma)}{4 \pi} c \Delta T',
\label{eq:phdis}
\end{eqnarray}
%%%%%%%%%%%%%%%%%%%%%%%
where we count photons escaped from both foreside ($\cos{\theta'} \geq 0$)
and backside ($\cos{\theta'} \leq 0$) of the shell with
the Lorentz transformation for $\theta$
%%%%%%%%%%%%%%%%%%%%%%%
\begin{eqnarray}
\cos{\theta'}=
\frac{\cos{\theta}-\beta_{\rm sh}}{1-\beta_{\rm sh} \cos{\theta}}.
\end{eqnarray}
%%%%%%%%%%%%%%%%%%%%%%%
For high latitude emissions with $\cos{\theta'} \leq 0$,
the shell thickness adds an extra ``time delay''
%%%%%%%%%%%%%%%%%%%%%%%
\begin{eqnarray}
\Delta t_{\rm bk}=(1+z)\frac{W' \Gamma}{c} \left(\cos{\theta}-\beta_{\rm sh} \right),
\end{eqnarray}
%%%%%%%%%%%%%%%%%%%%%%%
in Eq. (\ref{T-LT}).
The number of photons escaping from a surface element $\Delta S'$
%%%%%%%%%%%%%%%%%%%%%%%
\begin{eqnarray}
\frac{dN'(\varepsilon'_\gamma)}{d \Omega' dS'} \Delta S',
\end{eqnarray}
%%%%%%%%%%%%%%%%%%%%%%%
is transformed to the frame of the central engine with
%%%%%%%%%%%%%%%%%%%%%%%
\begin{eqnarray}
d S'=d S=2 \pi R^2 \sin{\theta} d \theta,
d \Omega'=\frac{d \Omega}{\Gamma^2 (1-\beta_{\rm sh} \cos{\theta})^2},
dN'=dN.
\end{eqnarray}
%%%%%%%%%%%%%%%%%%%%%%%
In each corresponding time-bin, the observer counts
the contribution from $\Delta S$ for the photon number 
per unit surface as
%%%%%%%%%%%%%%%%%%%%%%%
\begin{eqnarray}
\frac{d N_{\rm obs}}{dS_{\rm obs}}=\frac{1}{D_{\rm p}^2} \frac{dN}{d \Omega dS}
\Delta S,
\end{eqnarray}
%%%%%%%%%%%%%%%%%%%%%%%
where the proper distance $D_{\rm p}=D_{\rm L}/(1+z)$
is used, because we have already included the effects of the energy shift
and time-dilation due to cosmological redshift in eqs. (\ref{E-LT})
and (\ref{T-LT}).
Integrating over the surface with $0 \leq \theta \leq \theta_{\rm jet}$,
we obtain the total fluence for each time-bin.

As was calculated in \citet{mur07}, the photon absorption due to
extra galactic background light (EBL) is estimated by accumulating
the optical depth between the source and observer
with the EBL model of \citet{kne04} (best-fit).

\section{Results}
\label{sec:results}

The model parameters are five: the initial radius $R_0$, the bulk Lorentz factor 
$\Gamma$, the comoving magnetic field $B'$, the electron injection energy $E_{\rm e}$,
and the minimum electron energy $\varepsilon'_{\rm min}$.
While we can consider more complicated situations by evolving e.g., $B'$, $W'$, 
$\Gamma$, the electron injection rate, etc., with $R$, we adopt here for simplicity
the constant values of the parameters as mentioned in \S \ref{sec:model}.
The other fixed parameters are the jet opening angle $\theta_{\rm jet} = 5/\Gamma$,
electron injection index $p=2.5$,
and the parameter for the electron acceleration timescale $\xi=1$.

Since we make no assumption about the energy density of the ``thermal'' 
components, the conventional parameters of the energy fraction to the total
internal energy density, such as $\epsilon_{\rm e}$
and $\epsilon_B$ \citep[see the definitions in][]{mes06}, are not specified.
But assuming $\epsilon_{\rm e}=0.5$ with the electron energy density
$E_{\rm e}/(4 \pi R_0^3)$, hereafter we will take $\epsilon_B$ as a 
referential value for each parameter set.
The Thomson optical depth is another criterial parameter.
The electron number density in our simulations can be approximated as
%%%%%%%%%%%%%%%%%%%%%%%
\begin{eqnarray}
{n'}_{\rm e}=\frac{(p-2) E_{\rm e}}{(p-1)4 \pi \varepsilon'_{\rm min} R_0^3}.
\label{eqndash}
\end{eqnarray}
%%%%%%%%%%%%%%%%%%%%%%%
In all our simulations, the Thomson optical depth $\tau_{\rm T}={n'}_{\rm e}
\sigma_{\rm T} R_0/\Gamma$ is much smaller than unity.

\subsection{GRBs with ``Moderate'' Parameters}
\label{sec:moderate}

In this subsection we discuss three cases labeled Run1, Run2, and Run3,
whose parameters are indicated in the following figures \ref{fig:1}-\ref{fig:8}.
In Figure \ref{fig:1}, we show the evolution of
the spectral photon density in the shell frame for Run1.
The model parameter set is one to be considered as
fiducial values before the {\it Fermi} satellite was launched
(if $\epsilon_{\rm e}=0.5$, $\epsilon_B=0.0025$; $\tau_{\rm T}=0.03$).
As the electron injection proceeds from $R/R_0=1$, the photon
energy density gradually grows.
In this case the energy density of the magnetic field
is $\sim 4 \times 10^6~{\rm erg}~{\rm cm}^{-3}$, which
can be compared with the vertical axis in Figure \ref{fig:1}.
At an early stage (see $R/R_0=1.1$ in Figure \ref{fig:1})
synchrotron emission is the dominant cooling process for
injected electrons so that the spectral peak associated with $\varepsilon'_{\rm min}$
is seen at a few keV.
The relatively low photon density at this time leads to
a higher cut-off energy ($\sim 10$ MeV) due to $\gamma \gamma$-absorption.
As the photon energy density increases, the IC component gradually
becomes prominent (see $R/R_0=1.2$ and $1.5$), and
the cut-off energy due to $\gamma \gamma$-absorption decreases.
This late growth of the IC component has been
observed also in the simulations of \citet{bos09}.
The IC component makes the spectrum so hard that the synchrotron
peak gradually disappears, and instead,
$\gamma \gamma$-absorption produces a spectral peak
at a few hundreds keV.

Below the synchrotron peak energy at a few keV,
the synchrotron photons from cooled electrons
of $\varepsilon'_{\rm e}<\varepsilon'_{\rm min}$ make
a power-law spectrum with the photon index $\sim -1.5$.
The low-energy cut-off due to SSA
is clearly seen at $\sim 0.1$ eV.
The electron heating via SSA \citep{ghi88} makes a spectral bump
just above the cut-off energy.
At the end of the electron injection ($R/R_0=2$),
the superposition of the synchrotron and IC components
shows a hard power-law spectrum above the synchrotron peak energy.
Considering the Klein-Nishina effect,
the majority of seed photons for the IC process would be low-energy photons
at $\sim 1$ eV.
Above the cut-off energy at a few hundreds keV,
photon production and $\gamma \gamma$-absorption balance with each other,
and consequently generate a steep power-law spectrum.
After the electron injection stops, the shell expansion and photon escape
dilute the photon density (see the line of $R/R_0=2.5$).
In this stage, the photon production is practically terminated
so that our calculation based on the one-zone approximation
shows a sharp cut-off above the $\gamma \gamma$ cut-off energy.
The photon energy density in the shell frame
at $R/R_0=2$ is $1.4 \times 10^8~{\rm erg}~{\rm cm}^{-3}$,
which is close to the simple estimate
(note that $W'$ is constant) $E_{\rm e}/(4 \pi R^2 R_0)$ with
$R=2 R_0$, $2.0 \times 10^8~{\rm erg}~{\rm cm}^{-3}$.

\begin{figure}[htb!]
\centering
\epsscale{1.0}
\plotone{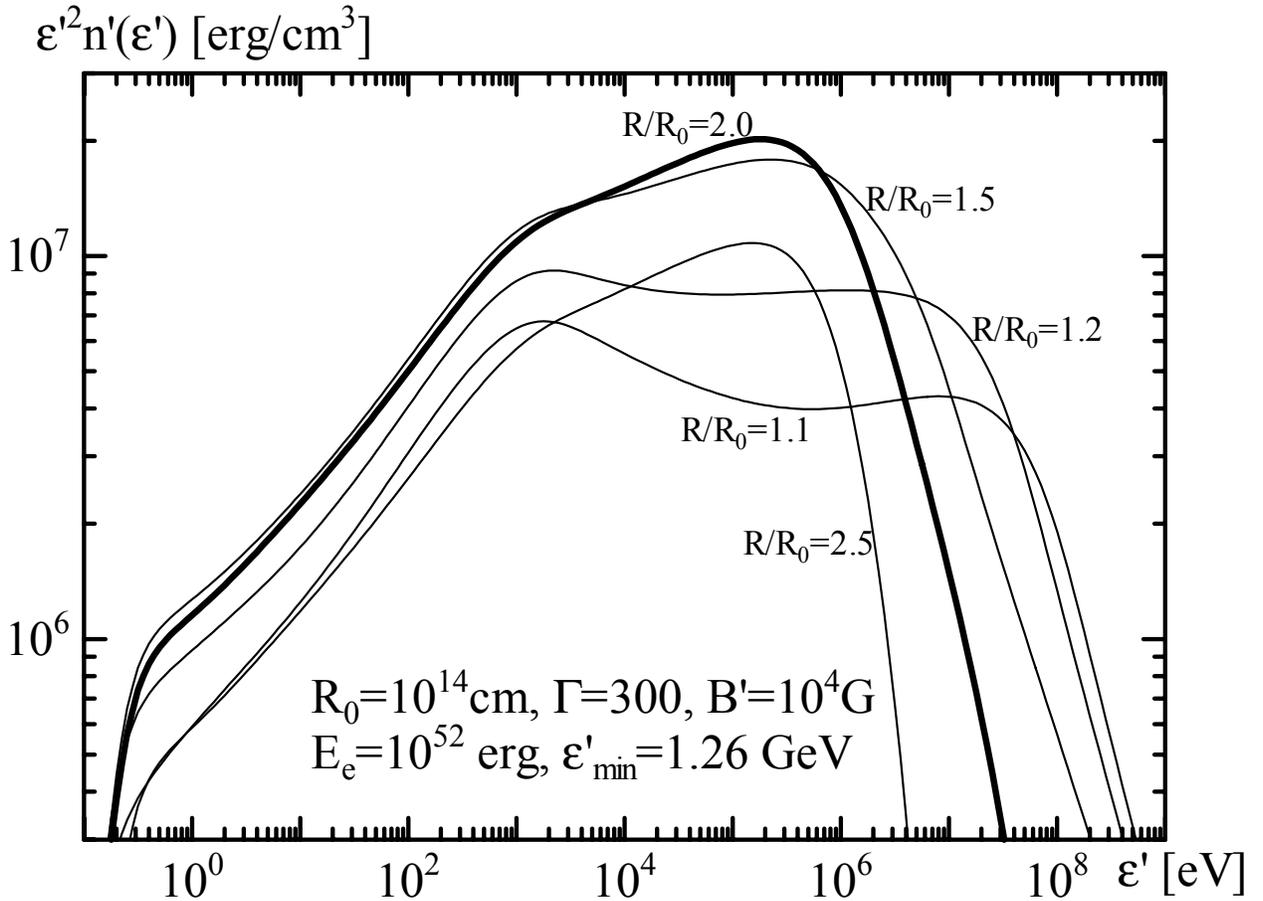}
\caption{Temporal evolution of spectral energy density of photons in the shell frame
for Run1.
The model parameters are denoted inside the figure.
The electron injection is ended at $R/R_0=2.0$ (thick line).
\label{fig:1}}
\end{figure}

\begin{figure}[htb!]
\centering
\epsscale{1.0}
\plotone{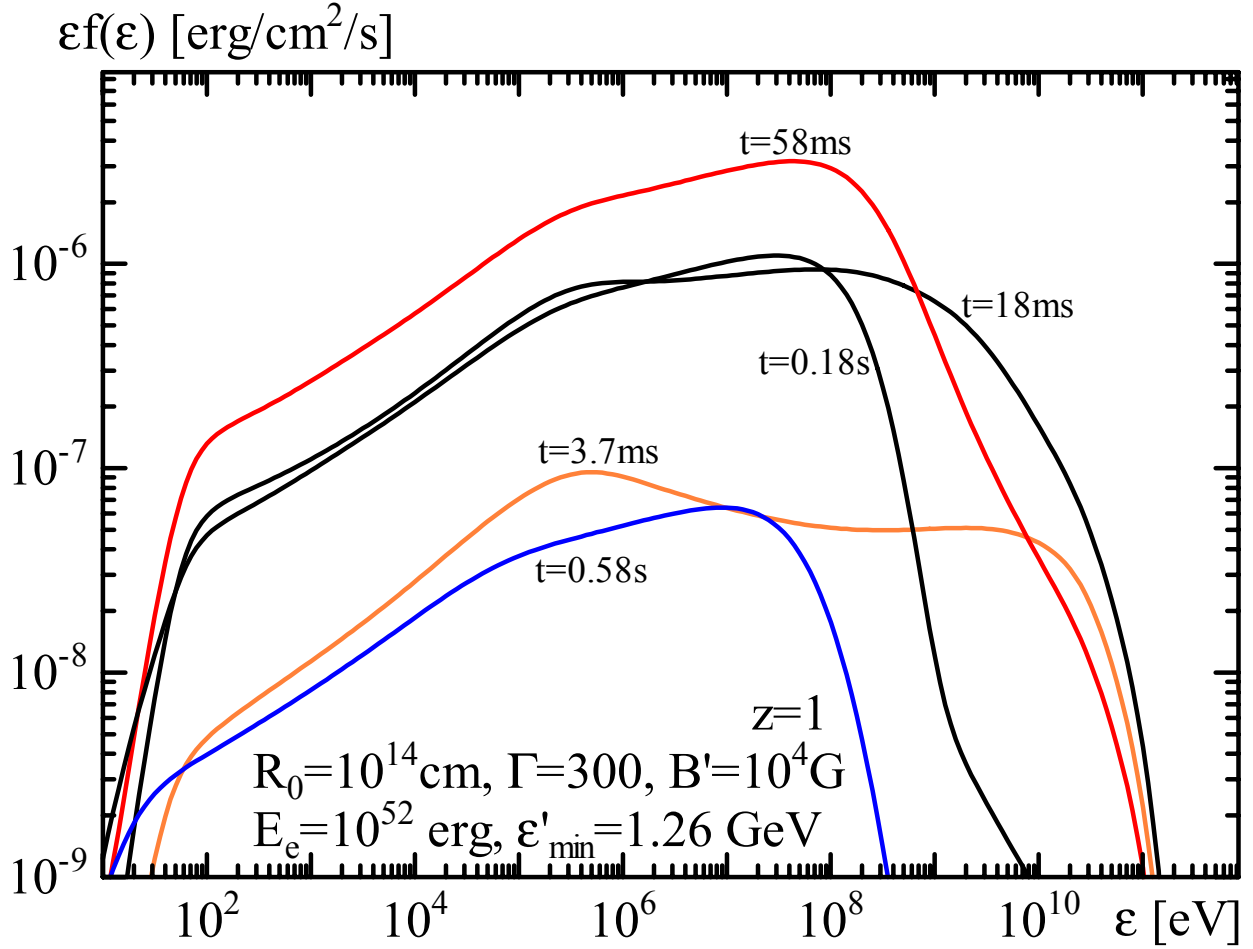}
\caption{Temporal evolution of the observable spectral photon flux
for the model described in Figure \ref{fig:1} (Run1).
The time-bins are the same as in Figure \ref{fig:3},
and the denoted times are the center for each time-bin.
\label{fig:2}}
\end{figure}

According to the method described in \S \ref{sec:LC}, we plot light curves
based on the spectral flux (the unit is
${\rm erg}~{\rm cm}^{-2}~{\rm s}^{-1}~{\rm eV}^{-1}$)
assuming the source redshift of $z=1$ in Figure \ref{fig:3}.
The time-bins are logarithmically divided (0.25 dex normalized
by $(1+z) R_0 \Gamma^{-2} c^{-1}$), and the flux
is averaged over each time-bin.
In the figure, the GeV, 100 MeV, and 100 keV light curves are plotted.
The keV light curve practically overlaps with the 100 keV lightcurve, and
the light curve shapes seem to reproduce well the typical FRED
\citep[fast rise, exponential decay;][]{fen96} shape,
which is frequently found in many GRBs.
Reflecting the temporal decrease of the $\gamma \gamma$ cut-off energy in 
the source, the GeV flux peaks earlier than the lower energy fluxes.
This corresponds to what one would get from an extrapolation of the 
``positive spectral lags'', observed in many of the pre-Fermi bursts \citep{nor96} 
(see also \S \ref{sec:lat}), to the GeV energy range.
This behavior is explicitly interpreted by the evolution of the photon 
spectrum shown in Figure \ref{fig:2}.
Before the peak time of the 1 keV-100 MeV fluxes ($\sim 58$ ms),
the spectrum seems to reflect the evolution of the spectrum in the source frame.
During the decaying phase, the spectrum gradually shifts to lower energy
owing to the emissions from higher latitudes of the shell (curvature effect).
We can see that GeV emission ceases faster than lower energy bands,
which explains the narrow pulse shape in the GeV light curve.
As shown in Figure \ref{fig:4}, however, the GeV fluence
is dim compared to the MeV-fluence in this parameter set.

\begin{figure}[htb!]
\centering
\epsscale{1.0}
\plotone{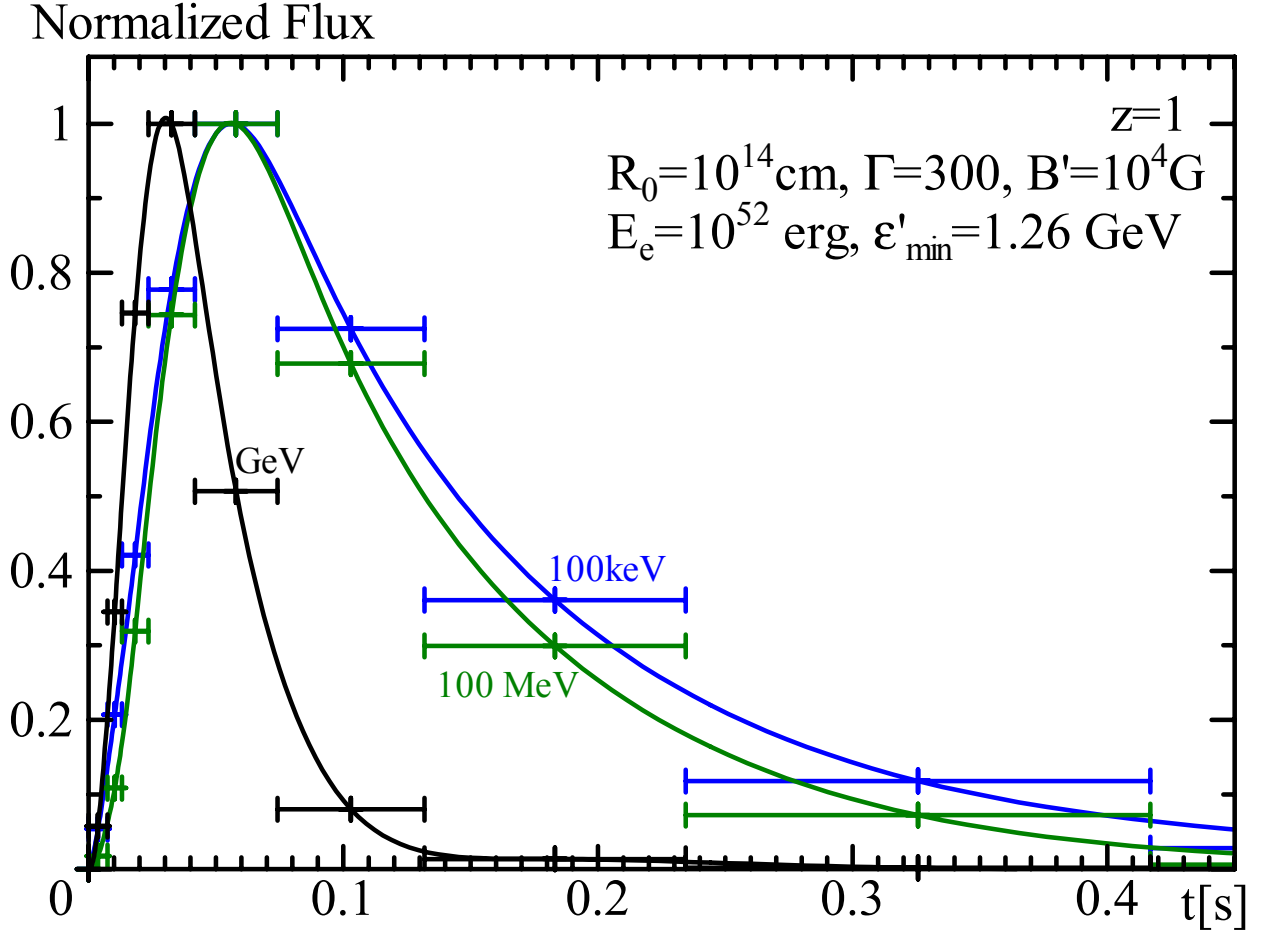}
\caption{Light curves for the model described in Figure \ref{fig:1} (Run1).
The plotted values are normalized by the maximum flux.
We smoothly join the fluxes averaged over respective time-bins
with solid curves for reference.
\label{fig:3}}
\end{figure}

\begin{figure}[htb!]
\centering
\epsscale{1.0}
\plotone{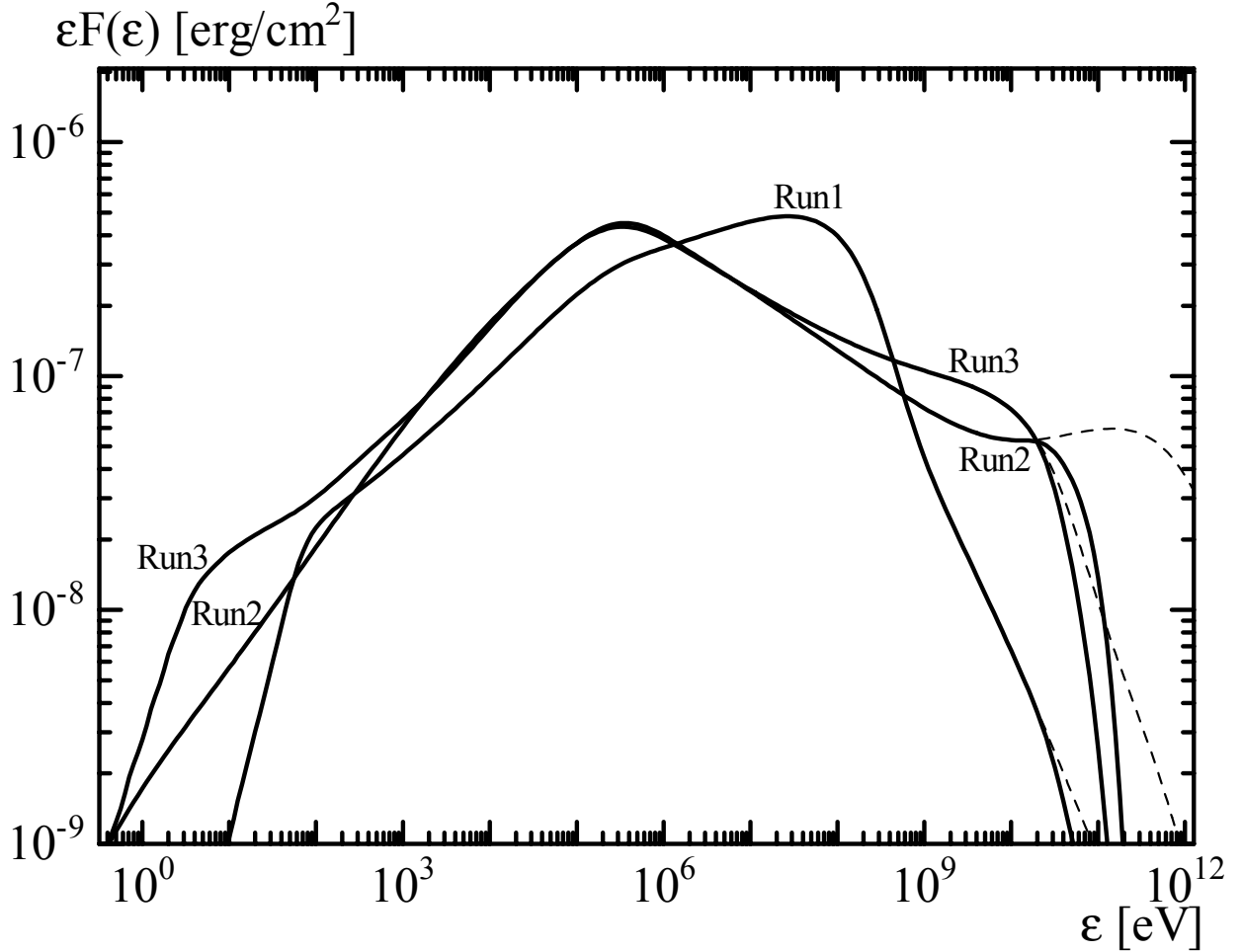}
\caption{Summarized observable fluences for Run1-Run3 with $z=1$.
The dashed lines denote the cases neglecting $\gamma \gamma$-absorption
due to EBL.
\label{fig:4}}
\end{figure}

We have shown the results for Run1 as an example that shows
a remarkably suggestive evolution of the spectral shape. However, 
the high-energy spectral index is harder than the typical observed ones.
Hence, we consider two different cases,
in which the internal $\gamma \gamma$-absorption for GeV photons
is not efficient, and the IC component is not so prominent;
Run2 with large $\Gamma$ and $R_0$  (Figures \ref{fig:5} and \ref{fig:6};
if $\epsilon_{\rm e}=0.5$, $\epsilon_B=0.06$; $\tau_{\rm T}=3 \times 10^{-6}$
),
and Run3 with moderate $\Gamma$ and
large magnetic field (Figures \ref{fig:7} and \ref{fig:8};
if $\epsilon_{\rm e}=0.5$, $\epsilon_B=0.025$; $\tau_{\rm T}=9 \times 10^{-5}$
).
The relatively stronger magnetic field than that in Run1
and the Klein-Nishina effect
make synchrotron radiation the dominant cooling process even though
$\epsilon_{\rm e}>\epsilon_B$.
In these cases % since the IC emission does not largely affect the
%electron cooling process,
the evolution of the fluxes (Figures \ref{fig:5} and \ref{fig:7})
are monotonic compared to Run1.

In Figure \ref{fig:5} the low-energy spectrum
below $\varepsilon_{\rm peak} \simeq 300$-400 keV
becomes soft with time.
The slow electron cooling due to
the low magnetic field in Run2
causes the slow evolution of the low-energy spectrum.
The typical energy of electrons emitting 1 eV synchrotron photons
is $\sim 8$ MeV. The synchrotron cooling time for such electrons
$\sim 550$ s is slightly longer than
$t'_{\rm inj} \sim 100$ s, although the IC cooling is marginally
important even after $t'=t'_{\rm inj}$.
This is the reason for the slow spectral evolution,
but note that $t'_{\rm cool}$ for electrons of $\varepsilon'_{\rm min}$
is much shorter than the dynamical timescale.
So almost all the energy injected in the shell
is released as photons even in this case.
For Run2 the SSA frequency is well below
the optical band so that
we can expect bright optical pulses as shown in Figure \ref{fig:4}.
While the light curves for keV-100 GeV regions are almost the same
as the 100 keV light curve in Figure \ref{fig:6},
the optical ($\sim 1$ eV) light curve shows a significant delay
relative to the higher energy bands.
This delay is a direct consequence of the gradual softening
of the low-energy spectrum.
The broader pulse profile of the 1 eV light curve compared to that of
the 100 keV light curve seen in Figure \ref{fig:6}
is due to the emission from higher latitudes.
As shown in Figure \ref{fig:4}, the intrinsic spectrum for Run2
extends as far as the 1 TeV range, where the IC component dominates,
but such photons are absorbed by EBL in our assumption of $z=1$.

\begin{figure}[htb!]
\centering
\epsscale{1.0}
\plotone{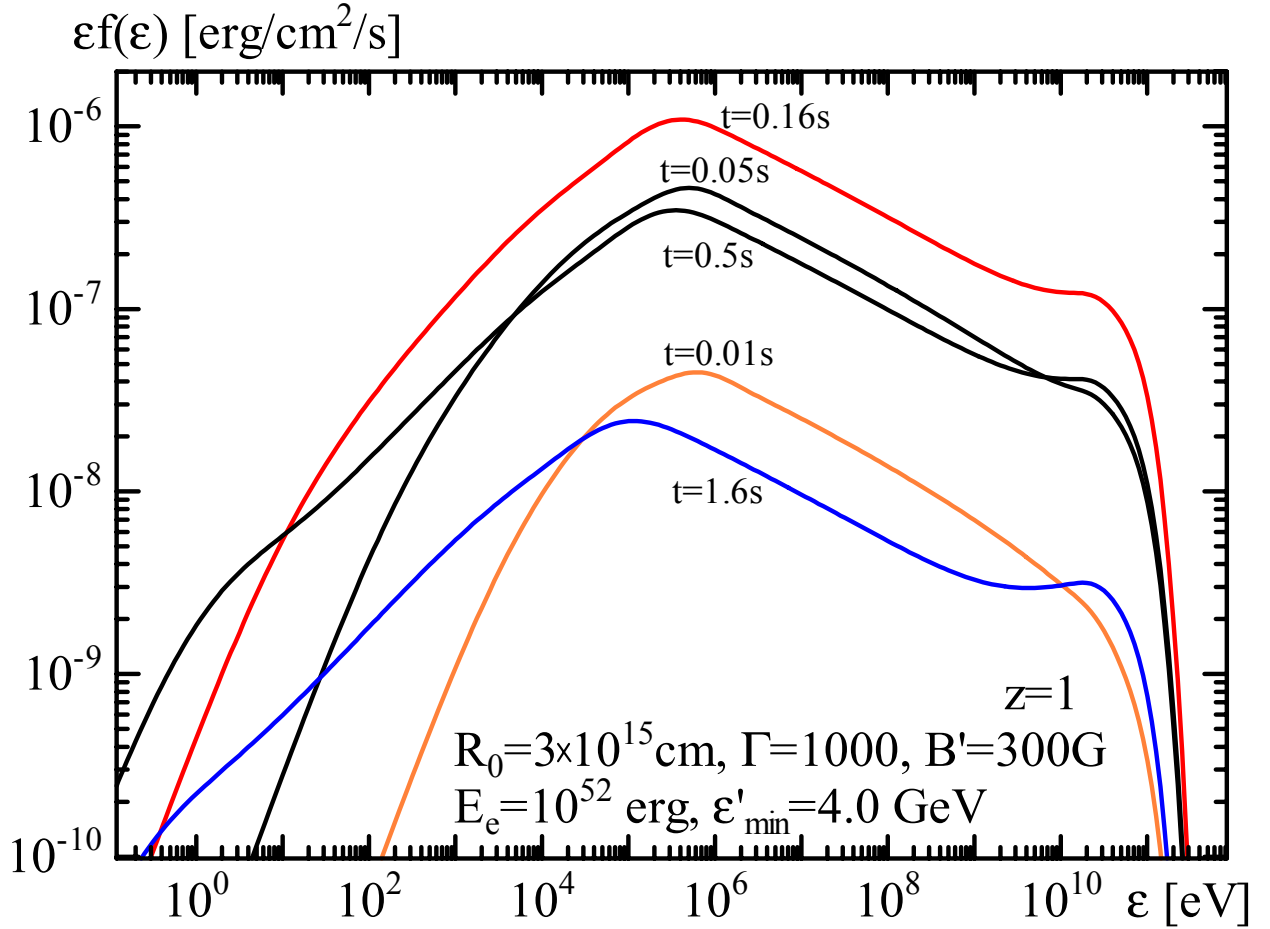}
\caption{Temporal evolution of the observable spectral photon flux
for the model Run2 (see also Figure \ref{fig:6}).
The time-bins are the same as in Figure \ref{fig:6},
and the denoted times are the center for each time-bin.
\label{fig:5}}
\end{figure}

\begin{figure}[htb!]
\centering
\epsscale{1.0}
\plotone{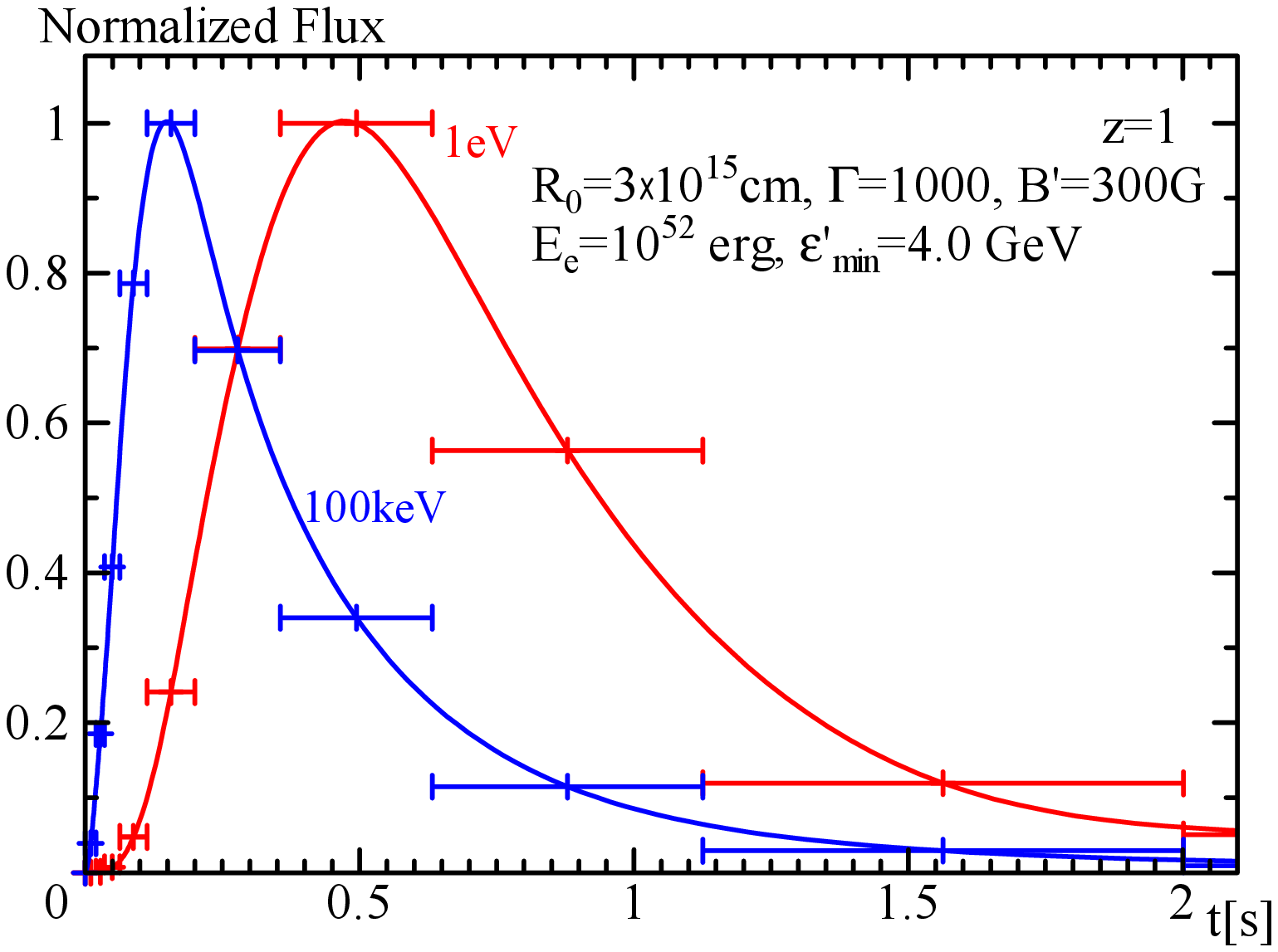}
\caption{The light curves for the model Run2, whose parameters
are denoted inside the figure.
The plotted values are normalized by the maximum flux.
We smoothly join the fluxes averaged over respective time-bins
with solid curves for reference.
\label{fig:6}}
\end{figure}

\begin{figure}[htb!]
\centering
\epsscale{1.0}
\plotone{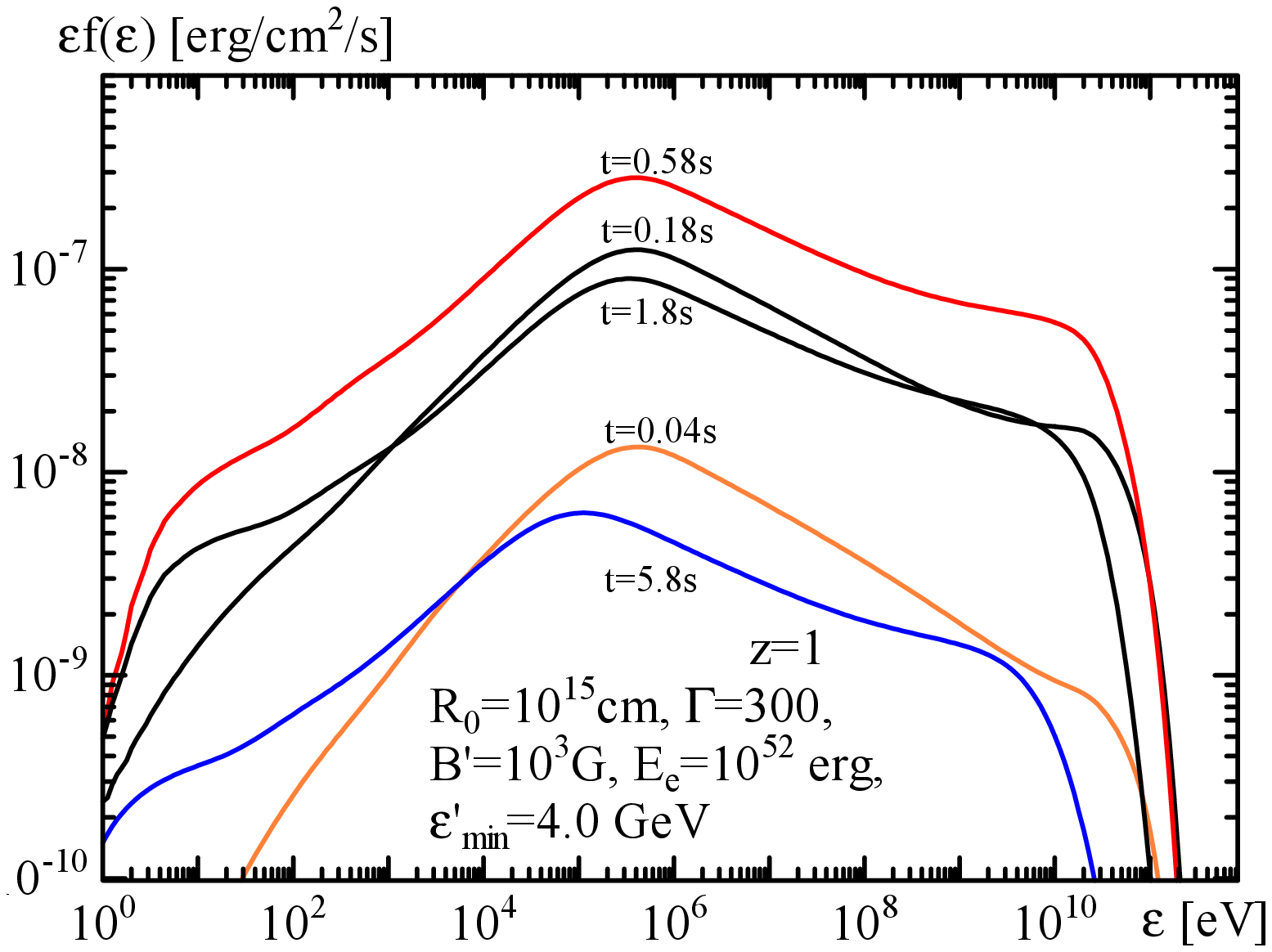}
\caption{Temporal evolution of the observable spectral photon flux
for the model Run3 (see also Figure \ref{fig:8}).
The time-bins are the same as in Figure \ref{fig:8},
and the denoted times are the center for each time-bin.
\label{fig:7}}
\end{figure}

\begin{figure}[htb!]
\centering
\epsscale{1.0}
\plotone{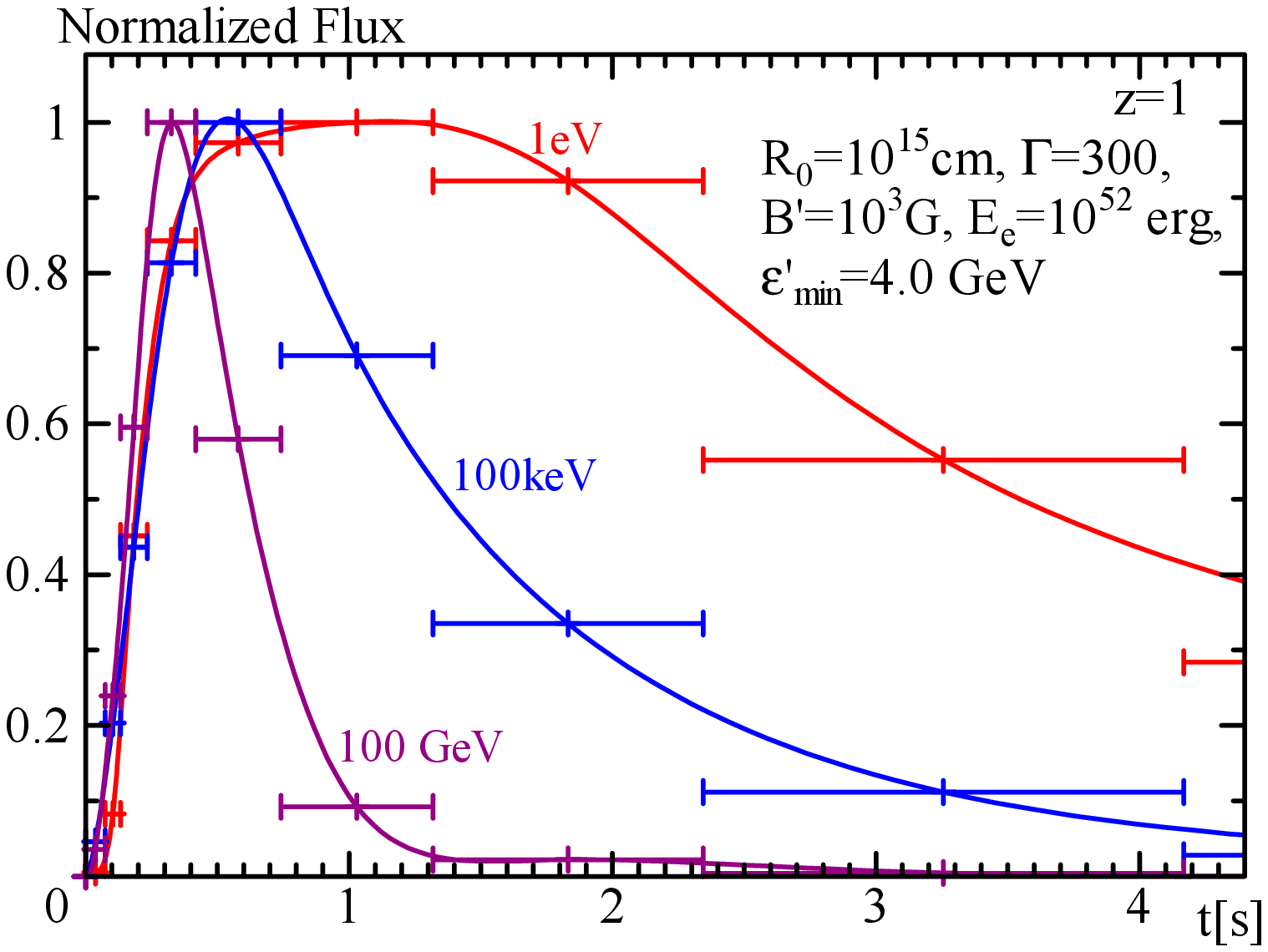}
\caption{Light curves for the model Run3, whose parameters
are denoted inside the figure.
The plotted values are normalized by the maximum flux.
We smoothly join the fluxes averaged over respective time-bins
with solid curves for reference.
\label{fig:8}}
\end{figure}

The time-integrated spectrum for the more compact case of Run3
is similar to that of Run2 (see Figure \ref{fig:4})
so that we can expect an optical pulse and
100 GeV photons possibly detectable with \v{C}erenkov detectors
such as MAGIC or CTA for those two cases.
The high-energy cut-off in the spectrum is mainly due to
the internal $\gamma \gamma$-absorption rather than due to the EBL,
unlike for the case of Run2.
The relatively fast electron cooling due to the high magnetic field in Run3
promptly generates a soft photon spectrum with a photon index $\sim -1.5$
below $\varepsilon_{\rm peak}$.
Thus, in this case no significant softening is found
during the pulse rise phase ($<0.58$s) unlikely Run2.
On the other hand, the smaller radius than that in the case of Run2
yields a slight evolution of the high energy
cut-off due to $\gamma \gamma$-absorption.
Therefore, the peak of the 100 GeV light curve appears earlier than
that of the lower energy bands,
while the light curves for keV-GeV regimes almost overlap with
the 100 keV lightcurve.
The heating of electrons due to SSA \citep{ghi88}
is so effective that a slight bump appears at a few eV
in the spectrum in Figure \ref{fig:4}.
The high-energy bump, prominent above 100 MeV compared to Run2,
may be due to secondary electron-positron pairs produced via
$\gamma \gamma$-absorption \citep{asa07}.

Above the $\gamma \gamma$ cut-off energy, the intrinsic
spectral shape for Run3 is well approximated by a power-law
rather than by an exponential cut-off.
This steep power-law shape is also seen in the spectrum of Run1.
If we take into account the spatial structure of the source,
a smoothly broken power-law would be expected as
a $\gamma \gamma$-absorption signature \citep{bar06,gra08}.
We see that, even with our one-zone approximation, the temporal 
evolution of the spectrum generates a broken power-law shape
in the time-integrated spectrum.
The poor statistics in the spectral data of GRB 090926A \citep{926A},
for which the spectral break is reported around 1.4 GeV,
makes it difficult to distinguish whether the spectral shape is one of 
a broken power-law or an exponential cut-off. Thus, future detections with
better statistics are needed of bright bursts with {\it Fermi} or with \v{C}erenkov 
telescopes, in order to determine more precisely the spectral shape above 
the cut-off energy.

Our results show that the curvature effect yields
a spectral lag between the optical and the keV-MeV bands. 
In the keV-MeV energy range, several authors have reported time delays
in the arrival time of lower energy photons relative to higher energy photons 
\citep[``positive lags'', e.g.,][]{nor96,wu00,hak04,che05,hak08,ari10}.
These spectral delays may be explained by the curvature effect \citep{iok01,she05,lu06}.
As \citet{lu06} discussed, the curvature effect predicts
that a smaller $\Gamma$ tends to yield a larger lag, i.e., softer photons
arrive increasingly later.
The optical delay in our results is consistent with this tendency,
the smaller $\Gamma$ in Run3 resulting in a broader 1 eV light curve
than that in Run2 (note that the optical delay in Run2 is mainly due to
the slower electron cooling).
However, the observed energy dependence of the lag does not always follow 
the prediction of the simple curvature effect \citep{zha07,ari10}.
We should also take into account the temporal evolution
of the spectral shape \citep{koc03,dai98,dai03,bos09}.
It would be interesting to fit the observed spectral lags by evolving
some of the model parameters, such as the magnetic field,
although it is not our purpose to discuss such effects in this paper.

The broad pulse profile in the optical band means that
the optical pulses tend to overlap each other.
The observed broader pulses and longer tail in the optical light curve
compared to the gamma-ray light curve in GRB 080319B \citep{rac08} is consistent
with this tendency, although an extra spectral component is required
to reproduce the optical flux in that burst.
Recalling that the slow electron cooling is a key to the
optical lag in Run2,
we emphasize that the spectral evolution should be investigated
not only in the GeV band but also in the optical band in order 
to probe the prompt emission mechanism.

\subsection{``Fermi-LAT'' Case}
\label{sec:lat}

The {\it Fermi} satellite detected GeV photons from
several very bright bursts ($E_{\rm iso}>10^{54}$ erg) such as
GRB 080916C \citep{916C}, GRB 090902B \citep{902B}, and
GRB 090926A \citep{926A}.
In such bursts the onset of the GeV emission is delayed with respect to 
the MeV emission, i.e., they have (in this energy range) negative lags.
In addition, some of them also have an extra spectral component above a
few GeV, distinct and additional to the usual Band function which dominates 
in the MeV energy range.
As \citet{cor10} discussed, the GeV emission may be due to IC
emission from internal dissipation regions.
In this subsection we choose a set of parameters, which may be relevant
for such bright GRBs accompanied by an additional GeV emission component, 
although we do not attempt to perform fits to the spectrum of any specific GRB.

\begin{figure}[htb!]
\centering
\epsscale{1.0}
\plotone{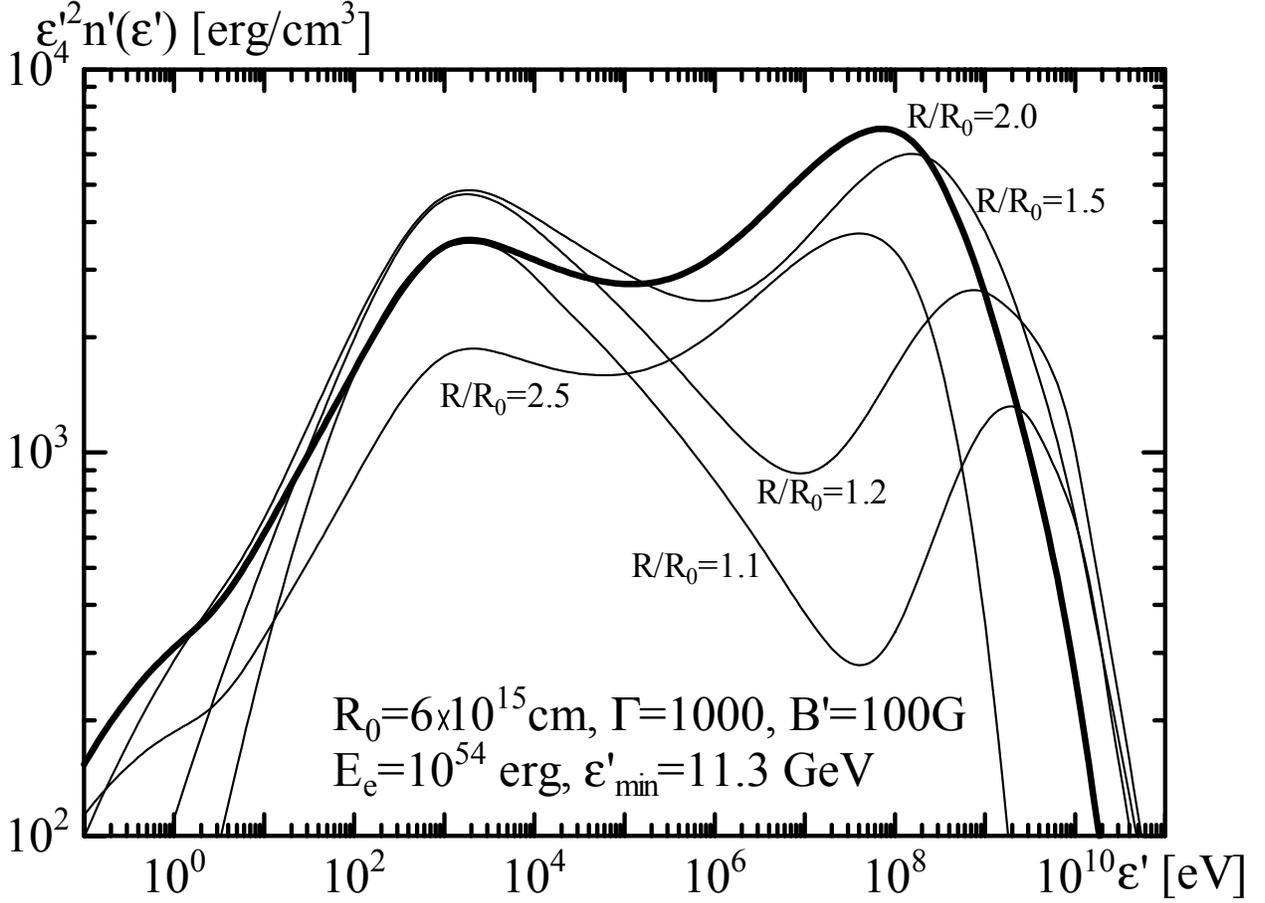}
\caption{Temporal evolution of the spectral energy density
of photons in the shell frame
for Run4.
The model parameters are denoted inside the figure.
The electron injection is ended at $R/R_0=2.0$ (thick line).
\label{fig:9}}
\end{figure}

In Figure \ref{fig:9}, we plot the evolution of the spectral photon density 
in the shell frame for Run4, 
in which we have adopted a very large injection energy ($E_{\rm e}=10^{54}$ erg)
and a high Lorentz factor ($\Gamma=1000$).
If we adopt $\epsilon_{\rm e}=0.5$, $B=100$ G corresponds
to $\epsilon_B=5.4 \times 10^{-4}$.
The Thomson optical depth is low enough, $\tau_{\rm T}=3 \times 10^{-5}$,
even for the large $E_{\rm e}$.

Initially the synchrotron component at $\sim$ keV dominates
the spectrum, but the IC component grows as electrons are continuously injected.
In the later stage the injected electrons cool mainly via IC rather than synchrotron,
so the synchrotron component starts to decay earlier than the IC component.
At the end of the electron injection ($R/R_0=2$)
the cooled electrons are still relativistic, because
the energy density of the magnetic field
is low ($\sim 400~{\rm erg}~{\rm cm}^{-3}$).
Equating the synchrotron cooling timescale and the dynamical timescale
$W'/c$, the energy of the cooled electrons is
%%%%%%%%%%%%%%%%%%%%%%%
\begin{eqnarray}
\varepsilon'_{\rm e,cool}&=&\frac{6 \pi m_{\rm e}^2 c^4 \Gamma}
{\sigma_{\rm T} B'^2 R_0} \nonumber \\
&\simeq& 200 {\rm MeV} \left( \frac{B'}{100~{\rm G}} \right)^{-2}
\left( \frac{\Gamma}{1000} \right)
\left( \frac{R_0}{6 \times 10^{15}~{\rm cm}} \right)^{-1}.
\end{eqnarray}
%%%%%%%%%%%%%%%%%%%%%%%
The contribution of IC cooling with the Klein-Nishina effect
can also be comparable to the above estimate,
as the obtained spectra show; the photon energy densities of the IC and
synchrotron components are comparable.
Actually, our numerical results indicate a spectral bump in the electron 
distribution around $100$ MeV at $R/R_0=2$.
Even after the end of electron injection, such cooled electrons
continue emitting synchrotron photons.
The cooling due to IC gradually becomes inefficient as
the seed photon density around keV energies decreases.
The typical synchrotron photon energy from
the cooled electrons is
%%%%%%%%%%%%%%%%%%%%%%%
\begin{eqnarray}
\varepsilon'_{\gamma,{\rm cool}}&=&\frac{3 \pi \hbar e B'}
{8 m_{\rm e} c} \gamma'^2_{\rm e,cool} \nonumber \\
&\simeq& 0.2 {\rm eV} \left( \frac{B'}{100~{\rm G}} \right)^{-3}
\left( \frac{\Gamma}{1000} \right)^2
\left( \frac{R_0}{6 \times 10^{15}~{\rm cm}} \right)^{-2}.
\end{eqnarray}
%%%%%%%%%%%%%%%%%%%%%%%
In the spectra for $R/R_0=2$ and $2.5$
we can see a spectral bump below $\sim 1$ eV,
which is attributed to the late synchrotron emission
from the cooled electrons.

\begin{figure}[htb!]
\centering
\epsscale{1.0}
\plotone{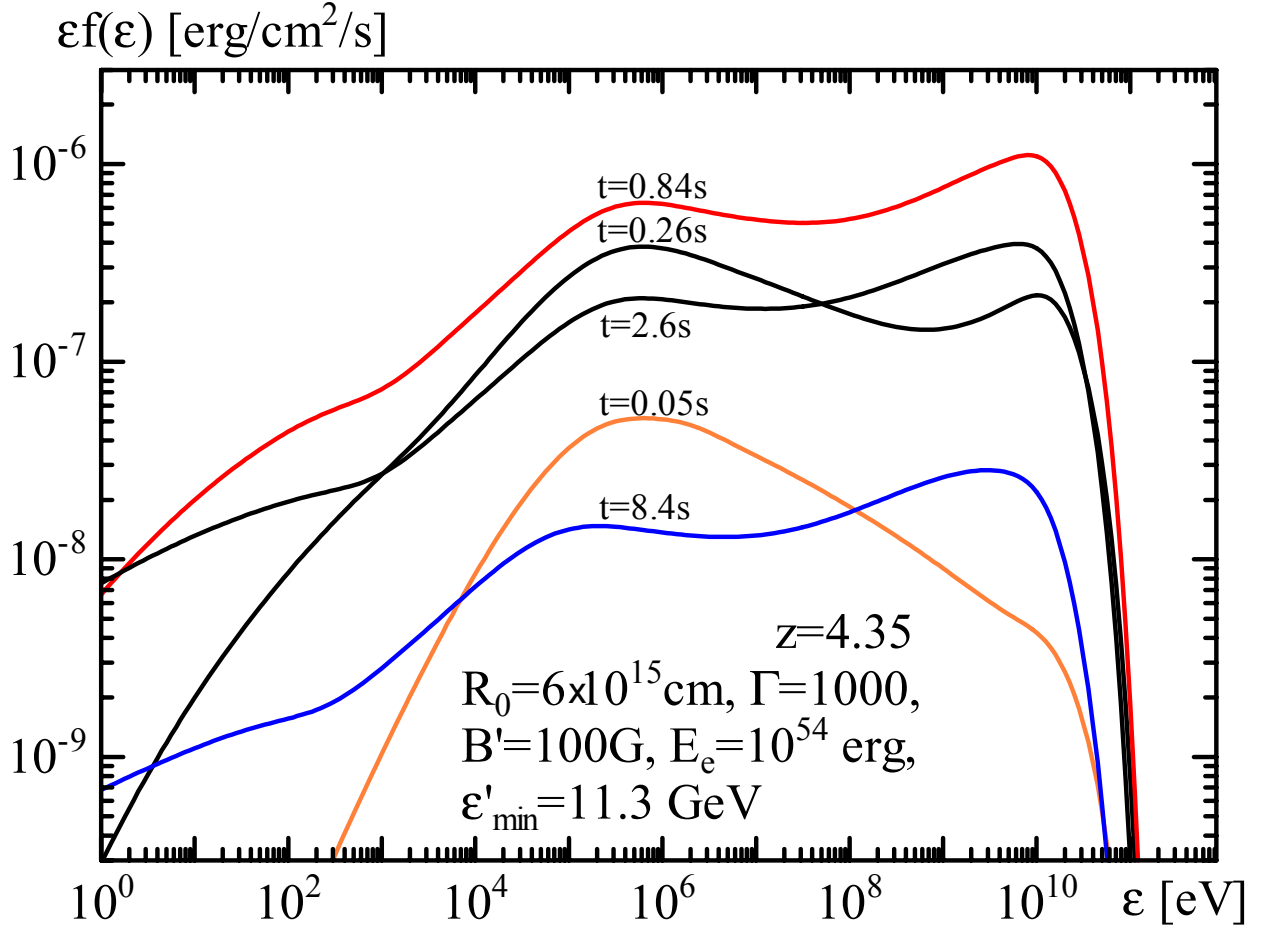}
\caption{Temporal evolution of the observable spectral photon flux
for the model Run4 with $z=4.35$.
The time-bins are the same as the logarithmic time-bins
in Figure \ref{fig:13},
and the denoted times are the center for each time-bin.
\label{fig:10}}
\end{figure}

Assuming a redshift $z=4.35$ such as that of as GRB 080916C, we show
the evolution of the spectrum in the observer frame for this Run4 case
in Figure \ref{fig:10}.
The IC component in the GeV-10 GeV range grows with time,
and the late synchrotron emission discussed above
produces a spectral break at keV $\sim \Gamma \varepsilon'_{\gamma,{\rm cool}}$.
The spectrum seems to evolve from the initial Band-function
to a power-law dominated shape in the later stages.
This spectral evolution is interestingly similar to that seen
in GRB 090510 \citep{510,510b}, although that burst is a short GRB.
The spectrum evolves to a power-law-like shape well
after the peak time of the pulse (typically $t=2.6$ s in Figure \ref{fig:10},
see also Figure \ref{fig:13}).
If many pulses overlap each other, a dominant pulse
may prevent the detection of the underlying power-law-like component.
Therefore, it is interesting that GRB 090510 is one of the short GRBs,
which have only one or two pulses in most cases.
In many {\it Fermi}-LAT GRBs, the spectra tend to evolve toward a
simple power-law spectrum extending as far as GeV in the later stage
of the prompt emissions.
Even for such long GRBs, the spectral evolution seen in our simulation
would be applicable,
because active pulses may gradually disappear in the later stage.
While such spectra may be explained by the early onset of the afterglow
\citep{ghi10,kum10}, a combination
of the late synchrotron emission and a decaying IC component
can also produce a power-law-like spectrum.

\begin{figure}[htb!]
\centering
\epsscale{1.0}
\plotone{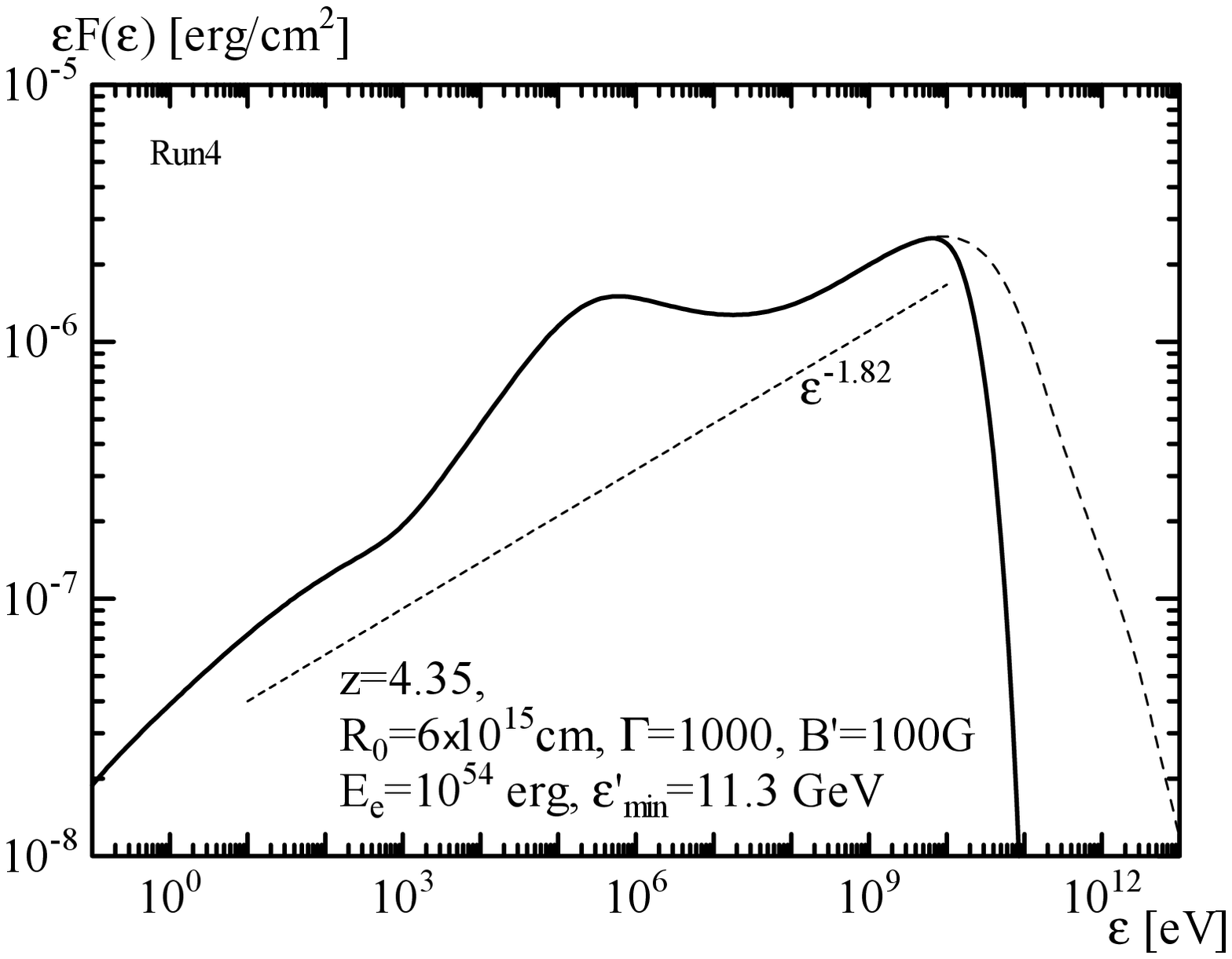}
\caption{Time-integrated observable flux for Run4 with $z=4.35$.
The dashed lines denote the cases neglecting $\gamma \gamma$-absorption
due to EBL.
\label{fig:11}}
\end{figure}

The time-integrated spectrum in Figure \ref{fig:11}
shows extra components below keV and above 100 MeV, comparable to those
seen in GRB 090510 \citep{510b} and GRB 090926A \citep{926A}.
To explain the observed extra components that dominates the emission
both below 10 keV and above 100 MeV,
a single contribution from an IC component is not sufficient.
However, the late synchrotron emission can assist in extending 
the extra component into the lower energy range, as shown in Figure \ref{fig:11}.
It seems that a single power-law component with a photon index $-1.82$,
which is close to the index $-1.79$ in GRB 090926A,
overlaps the usual Band function that dominates the emission
around MeV, despite the fact that
the two spectral excesses in the low and the high energy regions
have different physical origins.

\begin{figure}[htb!]
\centering
\epsscale{1.0}
\plotone{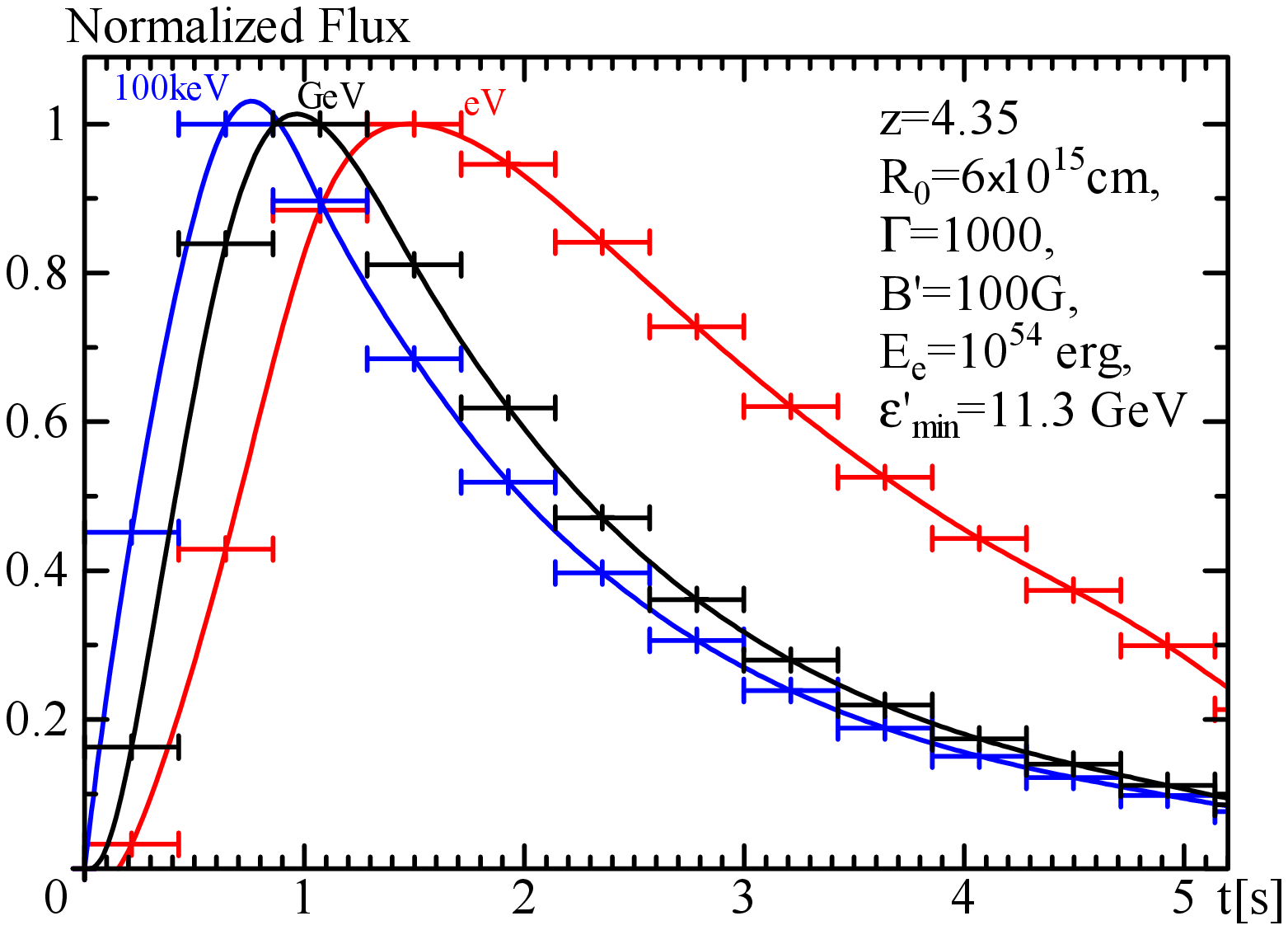}
\caption{Light curves for the model Run4 with $z=4.35$ based on the spectral flux.
The time-bins are linearly divided.
The plotted values are normalized by the maximum flux.
We smoothly join the fluxes averaged over respective time-bins
with solid curves for reference.
\label{fig:12}}
\end{figure}

In Figure \ref{fig:12} we plot the light curves based on the spectral flux
with linear time-bins, in order to distinguish the peak times.
The light curves show a delayed onset of the optical emission,
which is partially due to the late synchrotron emission.
The late growth of the IC emission makes
the growth of the early GeV light curve slower compared to the
100 keV emission \citep[see also][]{bos09},
the GeV peak time appearing delayed $\sim 0.4$ s
relative to the 100 keV peak time.
One needs however to take into account the low photon statistics
above 100 MeV in the GRB observations with {\it Fermi}. Thus,
in Figure \ref{fig:13} we plot the expected photon-count rates
integrating over the 100 MeV - 1 GeV and the $>1$ GeV ranges.
Here we have simply assumed constant effective areas of
3,000 ${\rm cm}^2$ and 7,000 ${\rm cm}^2$ for energies below and above
1 GeV, which mimic the effective are of the {\it Fermi}-LAT detector
\citep{ran09}.
While the count rates with logarithmic and linear binning
give us rather different impressions, the figure shows that we expect 
from this simple one-zone SSC model a detection of one photon with the 
{\it Fermi}-LAT, delayed at least $\sim 1$ s after the onset of the 100 keV 
or MeV emission. On the other hand, the observed delay of the GeV emission 
seen in GRB 080916C is $\sim 4$ s, which is substantially longer than 1 s.
Although the GeV delay due to the growth of the IC component would be within 
the approximate timescale of the keV-MeV pulse width, this delay appears able 
to explain only a fraction of the observed delayed onsets of the GeV emission
in long bursts.

\begin{figure}[htb!]
\centering
\epsscale{1.0}
\plotone{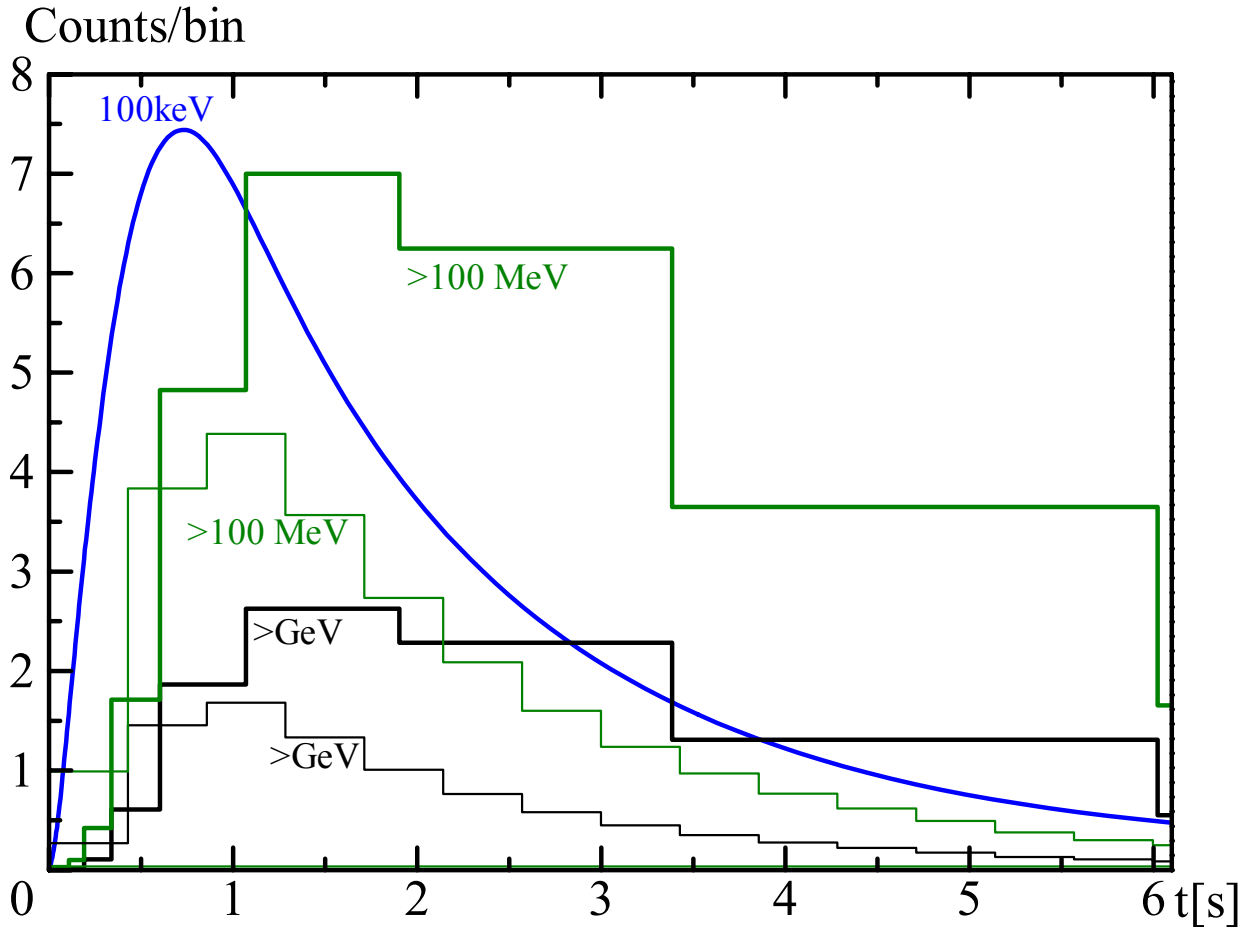}
\caption{Expected photon counts per each time-bin above
100 MeV and GeV for the model Run4 with $z=4.35$.
The thick (thin) histograms correspond to logarithmic (linear) time-bins.
For comparison we also plot the 100keV-lightcurve (arbitrary unit),
whose shape is the same as that in Fig. \ref{fig:12}.
\label{fig:13}}
\end{figure}

\subsection{External Photons}
\label{sec:external}

As discussed in the last subsection, the IC emission tends to be delayed 
relative to the main keV-MeV emission, but the longer delay than the pulse 
timescale of the MeV emission such as observed in GRB 080916C \citep{916C}
is not explained by this effect only. As a possible solution,
\citet{tom09,tom10} pointed out that the delayed GeV onset
can be reproduced by the introduction of external photons such as
the X-ray emission from the cocoon \citep[see also][]{li10,mur11}.
The geometrical configuration in those models, namely
the spatial separation between the source of the external photons
and the site of the internal shock or generic dissipation region can
cause the delayed arrival of the up-scattered external photons.
In particular, the photospheric emission \citep{pac86,mes00}
can provide the external seed photons for the IC emission from
electrons accelerated in regions outside the photosphere.
As suggested, e.g.,  for GRB 090902B \citep{ryd10,zha11},
many authors have discussed the photospheric emission
as the main MeV component.

In this subsection we assume that a quasi-steady emission
provides the main MeV component, emanating from an inner region,
while an outside dissipation region producing accelerated particles
upscatters these photons, as \citet{tom10} assumed. 
The quasi-steady emission may originate from the photospheric emission,
but we do not specify its origin here.
The model details are as follows:
The spectral shape of the quasi-steady emission is
a Band function \citep{ban93}, whose parameters are the luminosity $L_{\rm seed}$,
peak energy $\varepsilon_{\rm peak}$, 
and spectral indices $\alpha$ and $\beta$.
Here, the external emission is assumed to start at a time $\sim R_0/c$
before the electron injection.
In this Band photon field, accelerated electrons are injected in a
shell expanding from $R=R_0$
in the same way as we did in the previous subsections.
In the shell frame, the external photon field should be anisotropic,
because the Band component may be emitted (depending on the specific model)
from a radius $\ll R_0$, and its source may have a different bulk Lorentz 
factor than the $\Gamma$ of the shell.
Although our one-zone numerical code cannot fully include the effects of
such an anisotropy, we endeavor to take it into account partially as follows.
As long as the electron distribution in the shell frame is isotropic,
the total IC emissivity, even for anisotropic seed photons,
is written with the average intensity of the seed photons as
%%%%%%%%%%%%%%%%%%%%%%%
\begin{eqnarray}
J(\varepsilon_\gamma) \equiv \frac{U_{\rm seed}(\varepsilon_\gamma) c}{4 \pi}
\end{eqnarray}
%%%%%%%%%%%%%%%%%%%%%%%
\citep{aha81,bru00},
where $U_{\rm seed}(\varepsilon_\gamma)$ is the spectral energy
density of the seed photons.
However, reflecting the distribution of the seed photons,
the resultant emissivity also becomes anisotropic.
If the seed photons are emitted from a radius $\ll R_0$,
we can consider highly beamed seed photons that radially propagate
even in the shell frame.
In this case the anisotropy in the IC emissivity may be approximately included by 
adding an extra factor $(1-\cos{\theta'})$ in Eq. (\ref{eq:phdis}) in the Thomson 
limit \citep{wan06,fan08,tom09}.
This is because electrons moving in the same direction as the photon beam have
a smaller probability to interact with those photons, while conversely 
the head-on collisions between electrons and photons is much more efficient.
In this approximation the IC emission becomes maximum from the marginally 
high-latitude surface of the shell, rather than from the on-axis surface.
For a totally beamed photon field, its spectral shape in the shell frame is 
obtained by shifting the above Band function to a lower energy
by a factor of $(1+\beta_{\rm sh}) \Gamma$.
The average intensity of the Band component in the shell frame
is calculated with a normalization of the energy density
%%%%%%%%%%%%%%%%%%%%%%%
\begin{eqnarray}
\int d \varepsilon'_\gamma U'_{\rm seed}(\varepsilon'_\gamma)
=\frac{L_{\rm seed}}{4 \pi c R^2 (1+\beta_{\rm sh}) \Gamma^2}
\propto R^{-2},
\end{eqnarray}
%%%%%%%%%%%%%%%%%%%%%%%
for the totally beamed case.  This provides an extra factor $(1+\beta_{\rm sh})$, 
compared to the isotropic seed-photon case.

\begin{figure}[htb!]
\centering
\epsscale{1.0}
\plotone{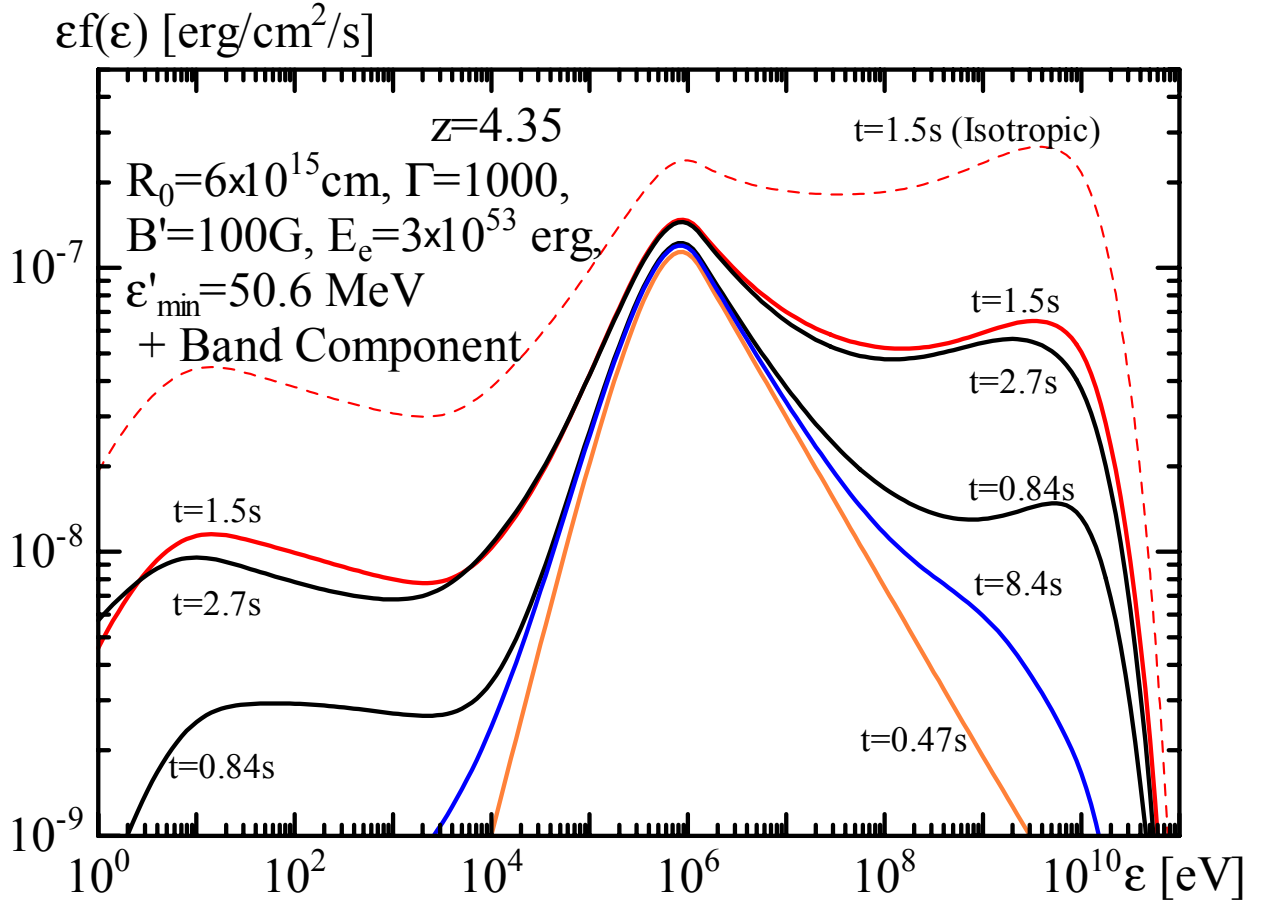}
\caption{Temporal evolution of the observable spectral photon flux
for the model Run5 including the external Band component with $z=4.35$.
The model parameters are denoted inside the figure.
The fluxes are averaged over the logarithmic time-bins
in Figure \ref{fig:15} and the denoted times are the center for each time-bin.
The solid lines are calculated based on the fully-beamed approximation,
while the thin dashed line represents the results in the isotropic approximation.
\label{fig:14}}
\end{figure}

The results shown below are obtained with the fully-beamed approximation mentioned
above. While expedient and useful, some of the weaknesses of such an approach 
are listed for completeness below.
The method in our code corrects for anisotropic effects all of the emissions from 
the shell, including synchrotron, which should be independent of
the anisotropy of the extra photon field.
Moreover, the seed photons for the IC include not only the external Band component
but also the photons emitted in the shell itself (SSC or secondary IC emissions).
Such secondary seed photons may soften the IC anisotropy.
The Klein-Nishina effects may also result in deviations from a simple anisotropic 
factor $(1-\cos{\theta'})$.
Even if the anisotropy of the emissivity in the shell is simply expressed by
this factor, the $\theta'$-dependence of the escape timescale should
yield the temporal evolution of the anisotropy of photon distribution.
Photons with small $|\cos{\theta'}|$ remain in the shell longer than
photons with $|\cos{\theta'}|\sim 1$. Thus, the angular distribution
should evolve even for the constant angular dependence of the emissivity.
Nevertheless, the one-zone and fully-beamed approximations force
the angular distribution to remain as $(1-\cos{\theta'})$.
Given the average photon energy $\overline{\varepsilon_\gamma}$
in the shell, our treatment results in
the average momentum of the escaped photons to become
~$-\overline{\varepsilon_\gamma}/(2c)$, although that of the photons
in the shell is ~$-\overline{\varepsilon_\gamma}/(3c)$.
Owing to the negative average momentum, the fluence for the observer
decreases by a factor two compared to the isotropic approximation.

The anisotropy of the Band component may affect $\gamma \gamma$-absorption as well
\citep{zou11,zhao11}.
The high-energy cut-off can be higher than that calculated here.
In this case too, however, the target photons against $\gamma \gamma$ are not only
the external photons. Since in any case our one-zone approximation is not suited 
for including the effect of the anisotropy in the absorption, we do not include 
these effects here.

\begin{figure}[htb!]
\centering
\epsscale{1.0}
\plotone{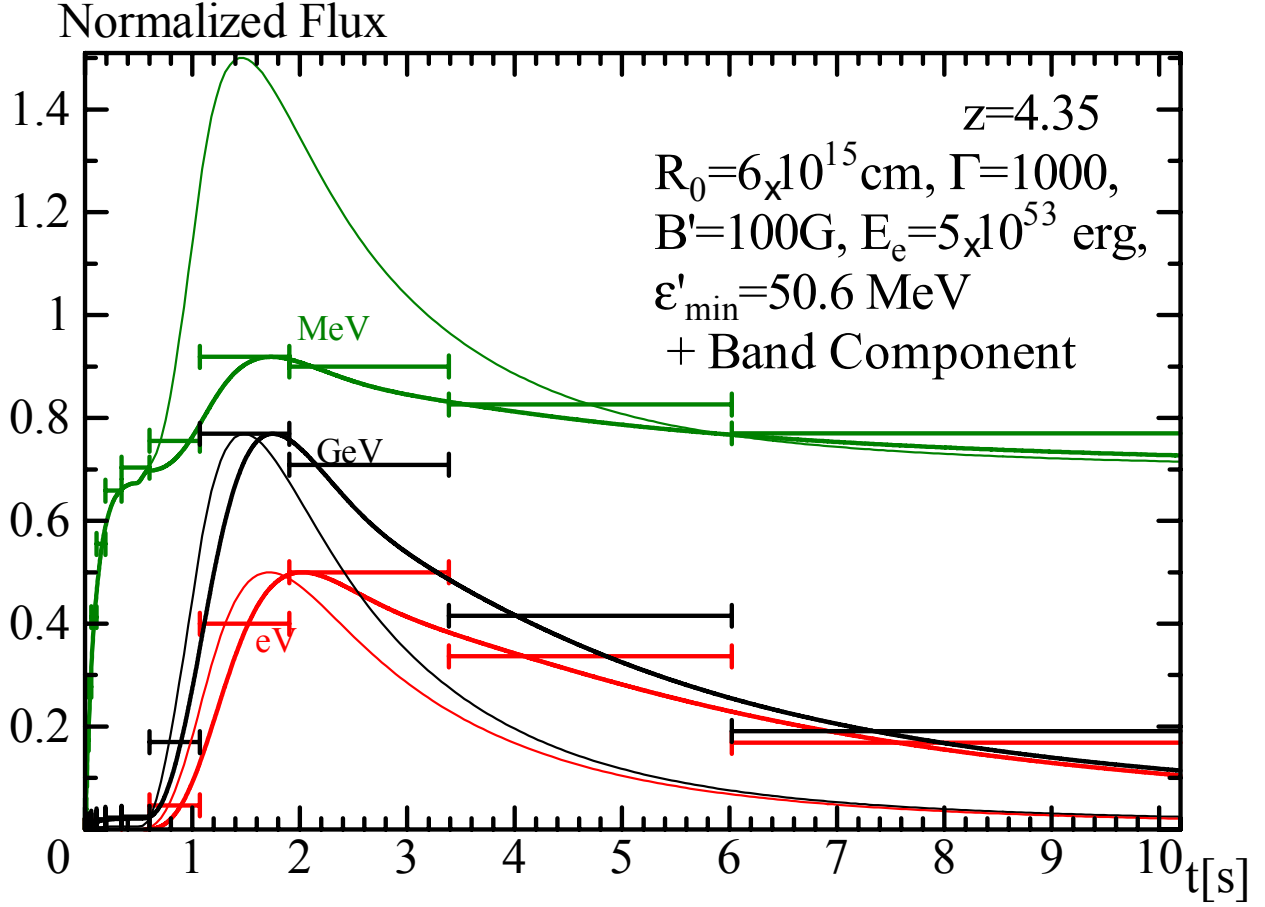}
\caption{Light curves for the model Run5 including the external Band component
with $z=4.35$ based on the spectral flux.
The normalizations are arbitrarily shifted to un-merge the lines.
The solid lines are plotted based on time-bins linearly divided ($0.0423$ s).
The fluxes averaged over logarithmic time-bins are also plotted.
The fully-beamed and isotropic approximations
are given by the thick and thin lines, respectively.
The peak fluxes for the thick and thin lines are aligned
in the GeV and eV light curves, while the MeV light curves are normalized
with the external photon flux.
\label{fig:15}}
\end{figure}

We adopt the parameters of the Band spectrum as
$L_{\rm seed}=10^{53}~{\rm erg}~{\rm s}^{-1}$ and
$\varepsilon_{\rm peak}\simeq 1$ MeV in the observer frame,
$\alpha=-0.6$, and $\beta=-2.6$
in the simulations of Run5 (Figures \ref{fig:14} and \ref{fig:15}).
The parameters for the internal shock
are similar to those in Run4 except for a small $\varepsilon'_{\rm min}$
(if $\epsilon_{\rm e}=0.5$, $\epsilon_B=0.0018$;
$\tau_{\rm T}=2 \times 10^{-3}$).
The small $\varepsilon'_{\rm min}$ leads to a large implied proton
energy.
The proton density ${n'}_{\rm p}$ would be the same as the electron density
estimated from Eq. (\ref{eqndash}). Even if we neglect the thermal
energy, the proton energy
$E_{\rm p} \sim \Gamma m_{\rm p} c^2 {n'}_{\rm p} V'
\sim 1.9 \times 10^{54}$ erg is fairly large.
In Figure \ref{fig:14} we show the spectral evolution for
the observer.
We see that the synchrotron and IC emission in the shell
contributes to a wide energy range, while the Band component
is prominent mainly in the MeV range.

In Figure \ref{fig:15} we show the corresponding light curves. We see that the
GeV and eV emissions originating from the (outer) shell are delayed 
$\sim (1+z)R_0/\Gamma^2/c \sim 1$ s relative to the MeV emission, which
is ``external'', i.e., it originates from a region which is at smaller radii
than the shell.
We should note that the start time of this external emission
can be arbitrary adjusted, depending on the model, so that the delay timescale
could be longer than shown by our results.
Since we assume a quasi-steady Band component,
the MeV emission continues even after the end of the GeV emission.
For comparison, we have also plotted in Figure \ref{fig:15} the light curves 
using the isotropic approximation for the photon emissions (thin lines).
Since the marginally high-latitude emission contributes the most to the flux
in the fully-beamed approximation, the light curves show larger delayed onsets
and longer tails
relative to those in the isotropic cases ($\sim 0.3$ s).
The Lorentz boost of the anisotropic emissions reduces
the fluence to half of that of the isotropic emission
owing to the negative average photon momentum in the shell frame.
Thus, those lower fluences and longer emission timescales
in the fully-beamed approximation result in
the lower fluxes shown in Figure \ref{fig:14}.
The origin of the eV emission is basically synchrotron, so the light curve
in the isotropic approximation (thin line) may be more appropriate for this.
Even for the IC emission, the actual flux evolution may be between the
cases represented by the fully-beamed and the simple isotropic approximation.

The internal shock produces the synchrotron and IC components
at $\sim 10$ eV and $\sim 10$ GeV, respectively.
The synchrotron peak appears able to explain the optical
extra component observed in GRB 080319B \citep{rac08}.
In the later stages, the seed photons for the IC process are not only
the Band component but also shell synchrotron photons,
so the IC emission contributes to 100 keV range as well as to the 10 GeV range.
As a result, the photon indices for the Band function below a few MeV
evolve to $\alpha \simeq -1.25$ and $\beta \simeq -2.35$ near the peak flux
($\alpha \simeq -1.52$ and $\beta \simeq -2.13$ for the isotropic approximation)
and return to the initial values later.
Similarly, the spectrum in GRB 080916C becomes flat
in $\varepsilon f(\varepsilon)$-plot, when the GeV emission brightens
\citep{916C}.
Thus, an internal shock with an external Band/photospheric component
appears generically able to explain the delayed onset of the GeV emission
and the spectral evolution observed in Fermi LAT GRBs.

While almost all the injected energy is released as photons
in the former cases of Run1-Run4, the small $\varepsilon'_{\rm min}$
in Run5 leads to a slower energy release.
Within the observer time $<6$ s, 79\% of the injected energy is released
in the isotropic approximation.
The integrated energy until the observer time $100(1+z)R_0/\Gamma^2/c \sim 100$ s
is 91\% of the injected energy
(we follow the photon emission from $R=R_0$ to $100 R_0$).
The rest of the energy is dissipated mostly via adiabatic cooling.

\section{Summary of the numerical results}
\label{sec:sum}

The results of our time-dependent simulations for the internal dissipation model
are summarized as follows.
\begin{itemize}
\item The temporal evolution of the photon energy distribution
inside the shell affects the resultant spectra seen by the observer.
Especially for weak magnetic fields, the evolution of the IC component
shows the limitation on the steady state approximation.
Time-resolved spectra are necessary in order to verify expected
features of the spectral evolution, such as the decrease of the cut-off 
energy due to $\gamma \gamma$-absorption, the growth of the IC component, 
and a spectral softening due to electron cooling.
\item Even in our one-zone approximation, the power-law spectrum
is produced above the $\gamma \gamma$ cut-off energy.
The spectral evolution and photon escape during the electron injection
are essential for this spectral shape.
\item We confirm that the light curves follow the FRED shape
\citep{fen96}. The pulse widths and peak times are affected
by not only the curvature effect but also the intrinsic evolution
of the energy distribution.
The decrease of the cut-off energy leads to an earlier GeV peak time,
and the gradual electron cooling causes the delayed onset
and broad pulse profile of the optical emission.
\item The possible deviation mechanisms from the simple Band function
(other than IC emission) are
synchrotron emissions from secondary electron-positron pairs,
electron heating via SSA, and late synchrotron emission
after the injection ends. The latter
two mechanisms yield a spectral bump
in the low energy portion, while the secondary pairs
enhance the synchrotron flux in the high energy range.
\item The retarded growth of the IC component can cause
the delayed onset of GeV emission.
This may be a partial reason for the delayed onsets
seen in the {\it Fermi}-LAT GRBs.
\item The combination of the IC and late synchrotron components
can produce the ``extra spectral component'' from keV to GeV
as seen in the {\it Fermi}-LAT GRBs.
This is a counter example to the claim that the IC emission
cannot explain the extra components in the {\it Fermi}-LAT GRBs.
\item When the IC and late synchrotron components are prominent,
a spectral evolution from the Band function to
a power-law dominated shape is seen.
This is similar to the spectral evolution observed in GRB 090510.
\item We confirm the essential features of external photon models
such as proposed by \citet{tom09,tom10}.
The IC emission using external Band photons coming from inside regions
can reproduce the delayed onset of GeV emission.
The effect of the anisotropic emissivity adds an extra factor
for the delay timescale.
The evolution of the spectral indices is similar to those
seen in GRB 080916C.
The long-lasting tail of GeV emission in this model is
also an interesting feature for the {\it Fermi}-LAT GRBs.
\end{itemize}

\section{Discussion}
\label{sec:disc}

Internal shocks, or more generally internal dissipation, have been considered 
as the main source to explain the GRB prompt emission \citep{pir05,mes06}, but several
problems, such as the narrow distribution of the peak energy $\varepsilon_{\rm peak}$
\citep{pre00} etc., have led to consideration of photosphere models \citep{pac86,mes00},
in which most of the outflow energy is released as thermal photons
from the photosphere.
More recently, dissipative photosphere models have been invoked to
explain the spectra and high efficiency in the prompt emission
\citep{gia07,iok07,pee08,ryd10,bel10,zha11,vur11}.
However, internal dissipation regions located outside the photosphere,
including the classical internal shock model,
are still attractive, as summarized in \S \ref{sec:sum}.
The emission from accelerated electrons in the dissipation region
provides a natural mechanism for emitting GeV photons
as detected with {\it Fermi}-LAT
\citep[as for the dissipative photosphere models,
see, e.g.,][]{bel10,vur11}.
Also for the light curve shape, the photospheric radius
may be much smaller than $c t_{\rm var} \Gamma^2$ ($t_{\rm var}$
is the variability timescale), so that the photosphere model
must attribute the broad pulse shape to
the long-duration activity of the central engine.
On the other hand, the internal dissipation models can easily
reproduce the FRED shape \citep{kob07} and the spectral lags \citep{she05}.
Although some fraction of GRBs such as GRB 090902B
may be explained with a simple photospheric component \citep{zha11},
the internal dissipation in the outflows can play an important role
not only inside the photosphere but also outside.

In this paper we test two models for the extra-spectral components
detected with {\it Fermi}-LAT; the first is the simple IC emission
from accelerated electrons and the second is the up-scattering of external photons.
Both models require an internal dissipation region in the outflow,
where electron acceleration occurs.
The possible delay time in the simple IC model is within the pulse
width, so in this model one needs some additional reason to explain the 
$\sim 4$ s delay in GRB 080916C.
However the external photon model has the additional advantage of 
being able to reproduce such delayed onsets of the GeV emission.
We note that external shock models with a high bulk-Lorentz factor
\citep{ghi10,kum10} can naturally explain the delayed onset of GeV emission, 
but as \citet{Max11} have pointed out, the early-phase energy output estimated 
from the GBM data are not sufficient to explain the GeV fluxes in the prompt phase
\citep[see also,][]{he11}.
The variability of the GeV emission needs to be tested with
much high photon statistics in order to verify whether their origin
is internal or external.

The cascade emissions initiated by hadronic interactions
\citep[e.g.,][]{boe98,gup07,asa07,asa09} are also one of the 
interesting options for explaining the GeV photon emission, 
related to the potential role of GRB as sources of
ultra high-energy cosmic rays (UHECRs) \citep{wax95,vie95}
and neutrino emission \citep{wax97,gue01,mur06}.
In particular, the flat spectrum of the extra component in GRB 090902B
is well explained by the hadronic model with a fiducial parameter set
\citep{asa10}.  However, that GRB may be an exceptional case;
the detection of GeV photons generally requires a large bulk Lorentz factor
and large emission radius to avoid the internal $\gamma \gamma$-absorption.
Such situations reduce the efficiency of
the photomeson production. As \citet{asa09b} showed,
a much larger proton luminosity than gamma-ray luminosity is required
to produce the extra spectral component in GRB 090510.
Although the high proton loading of the hadronic models
is favorable for the GRB-UHECR scenario, the leptonic models considered 
in this paper are energetically less demanding and appear to provide a 
broader agreement with the observed spectra and time delays.
The spectral evolution in the GeV-TeV range could provide essential clues
for distinguishing the leptonic and hadronic models.
This paper is thus both a test of purely leptonic GRB models based on SSC and
EIC schemes, and also a first step toward developing 
a time-dependent code to simulate emissions including the hadronic processes.
We plan to compare the evolutions of the GeV-TeV spectrum
for both the leptonic and hadronic models in near future.

Concerning the classical internal shock model, in this paper we have not
focused on the various problems mentioned with these, which have served as
a motivation for photosphere models. We note that the minimum energy of 
electrons $\varepsilon'_{\rm min}$, (conventionally expressed by the energy 
and number fractions of accelerated electrons) is treated as a free parameter 
in our simulations. Thus, our results do not address questions regarding the 
narrow distribution of $\varepsilon_{\rm peak}$ or the empirical 
$\varepsilon_{\rm peak}$-correlations such as the Yonetoku relation \citep{yon04}.
Similarly, the open problem of the low-energy spectral index $\alpha$
is not addressed here.
In our results in \S \ref{sec:moderate} and \ref{sec:lat}
the shorter cooling time than the dynamical timescale makes $\alpha \simeq -1.5$,
while the typical observed value is $\alpha=-1.0$ \citep{pre00}.
To solve the $\alpha$-problem, several mechanisms have been proposed, such as
the Klein-Nishina effect on SSC \citep{der01,bos09,nak09,wan09}
and the decay of magnetic fields \citep{pee06}, as well as the effect of
the superposition of different optical depth and temperature regions in 
photospheric models, e.g., \citet{bel10} and \citet{vur11}.
Other issues related to the magnetic geometry and expansion dynamics
or to nuclear collisions effects \citep[e.g.,][]{mes11} can affect the
location of the photosphere or provide additional delay mechanisms.
Also, in the dissipation region, we can expect magnetic turbulence
due to the Kelvin-Helmholtz instability \citep{zha09}
or the Richtmyer-Meshkov instability \citep{miz10,ino11}.
Such highly disturbed long-wave magnetic fields may interact with
high energy electrons.  The electron heating due to the turbulence may 
harden the spectrum \citep{asa09c}. Such effects expected in the dissipation 
region should be considered.

The code discussed in this paper will be further developed to treat hadronic 
processes as well. In the near future, we plan to carry out simulations for 
various situations involving dissipative photospheres and internal or external 
dissipation or shock regions.  In order to test with these a realistic picture
of GRB, such calculations will need to adopt a wide range parameter sets for the
magnetic field, bulk Lorentz factor and baryon loading \citep[see e.g.,][]{iok10,iok11}.
Moreover, the code will be useful for simulating emissions
of other high-energy sources, such as active galactic nuclei,
supernova remnant, and clusters of galaxies.

\begin{acknowledgments}
First we thank the anonymous referee.
This study is partially supported by Grants-in-Aid for Scientific Research
No.22740117 from the Ministry of Education,
Culture, Sports, Science and Technology (MEXT) of Japan,
NASA NNX08AL40G, and NSF PHY-0757155.
\end{acknowledgments}


\begin{thebibliography}{}

\bibitem[Abdo et al. (2009a)]{510}
Abdo, A. A. et al. 2009, \nat, 462, 331
\bibitem[Abdo et al. (2009b)]{902B}
Abdo, A. A. et al. 2009, \apj, 706 L138
\bibitem[Abdo et al. (2009c)]{916C}
Abdo, A. A. et al., 2009, Science, 323, 1688
\bibitem[Ackermann et al.(2011)]{926A}
Ackermann, M. et al.,  2011, \apj, 729, 114
\bibitem[Ackermann et al.(2010)]{510b}
Ackermann, M. et al.,  2010, \apj, 716, 1178
\bibitem[Aharonian \& Atoyan (1981)]{aha81}
Aharonian, F. A., \& Atoyan, A. M., 1981, \apss, 79, 321
\bibitem[Arimoto et al. (2010)]{ari10}
Arimoto, M. et al., 2010, \pasj, 62, 487
\bibitem[Asano (2005)]{asa05}
Asano, K. 2005, \apj, 623, 967
\bibitem[Asano et al. (2009a)]{asa09b}
Asano, K., Guiriec, S., \& M\'esz\'aros, P. 2009, \apj, 705  L191
\bibitem[Asano \& Inoue (2007)]{asa07}
Asano, K., \& Inoue, S. 2007, \apj, 671, 645
\bibitem[Asano et al. (2009b)]{asa09}
Asano, K., Inoue, S., \& M\'esz\'aros, P. 2009, \apj, 699, 953
\bibitem[Asano et al. (2010)]{asa10}
Asano, K., Inoue, S., \& M\'esz\'aros, P. 2010, \apj, 725, L121
\bibitem[Asano \& Nagataki (2006)]{asa06}
Asano, K., \& Nagataki, S. 2006, \apjl, 640, L9
\bibitem[Asano \& Terasawa (2009)]{asa09c}
Asano, K., \& Terasawa, S. 2009, \apj, 705, 1714
\bibitem[Band et al. (1993)]{ban93}
Band, D. et al. 1993, \apj, 413, 281
\bibitem[Baring (2006)]{bar06}
Baring, M. G., 2006, \apj, 650, 1004
\bibitem[Belmont et al. (2008)]{bel08}
Belmont, R., Malzac, J., \& Marcowith, A. 2008, \aap, 491, 617
\bibitem[Beloborodov (2005)]{bel05}
Beloborodov, A. M., 2005, \apj, 618, L13
\bibitem[Beloborodov(2010)]{bel10} Beloborodov, A. M. 2010, MNRAS, 407, 1033
\bibitem[Berestetskii et al.(1982)]{ber82}
Berestetskii, V. B., Lifshitz, E. M., \& Pitaevskii, L. P. 1982,
Quantum Electrodynamics (New York: Pergamon), 371
\bibitem[B\"ottcher \& Dermer (1998)]{boe98}
B\"ottcher, M., \& Dermer, C. D. 1998, \apj, 499, L131
\bibitem[Bo\v{s}njak et al. (2009)]{bos09}
Bo\v{s}njak, \v{Z}, Daigne, F., \& Dubus, G. 2009, \aap, 498, 677
\bibitem[Brunetti (2000)]{bru00}
Brunetti, G., 2000, Astropart. Phys., 13, 107
\bibitem[Chen et al.(2005)]{che05}
Chen, L. et al.,  2005, \apj, 619, 983
\bibitem[Corsi et al.(2010)]{cor10}
Corsi, A. Guetta, D., \& Piro, L.  2010, \aap, 524, 92
\bibitem[Daigne et al. (2011)]{dai11}
Daigne, F., Bo\v{s}njak, \v{Z}.,  \& Dubus, G. 2011, \aap, 526, 110
\bibitem[Daigne \& Mochkovitch (1998)]{dai98}
Daigne, F., \& Mochkovitch, R. 1998, \mnras, 296, 275
\bibitem[Daigne \& Mochkovitch (2003)]{dai03}
Daigne, F., \& Mochkovitch, R. 2003, \mnras, 342, 587
\bibitem[Derishev et al. (2001)]{der01}
Derishev, E. V., Kocharovsky, V. V., \& Kocharovsky, V., VI. 2001, \aap, 372, 1071
\bibitem[Fan et al.(2008)]{fan08}
Fan, Y.-Z., Piran, T., Narayan, R., \& Wei, D.-M. 2008, \mnras, 384, 1483
\bibitem[Fenimore et al.(1996)]{fen96}
Fenimore, E. E., Madras, C. M., \& Nayakshin, S. 1996, \apj, 473, 998
\bibitem[Ghisellini et al. (1988)]{ghi88}
Ghisellini, G., Guilbert, P. \& Svensson, R. 1988, \apj, 334, L5
\bibitem[Ghisellini et al. (1998)]{ghi98}
Ghisellini, G., Haardt, F., \& Svensson, R. 1998, \mnras, 297, 348
\bibitem[Ghisellini \& Svensson (1991)]{ghi91}
Ghisellini, G., \& Svensson, R. 1991, \mnras, 252, 313
\bibitem[Ghisellini et al. (2010)]{ghi10}
Ghisellini, G. at al. 2010, \mnras, 403, 926
\bibitem[Giannios \& Spruit (2007)]{gia07}
Giannious, D., \& Spruit, H. C. 2007, \aap, 469, 1
\bibitem[Gonz\'alez et al. (2003)]{gon03}
Gonz\'alez, M. M., Dingus, B. L., Kaneko, Y.,
Preece, R. D., Dermer, C. D., \& Briggs, M. S. 2003, \nat, 424, 749
\bibitem[Granot et al. (2008)]{gra08}
Granot, J., Cohen-Tanugi, J. \& do Couto e Silva, E. 2008, \apj, 677, 92
\bibitem[Guetta \& Granot (2003)]{gue03}
Guetta, D., \& Granot, J. 2003, \apj, 585, 885
\bibitem[Guetta et al.(2001)]{gue01}
Guetta, D., Spada, M., \& Waxman, E. 2001b, \apj, 559, 101
\bibitem[Gupta \& Zhang (2007)]{gup07}
Gupta, N., \& Zhang, B., 2007, \mnras, 380, 78
\bibitem[Hakkila \& Giblin (2004)]{hak04}
Hakkila, J., \& Giblin, T. W. 2004, \apj, 610, 361
\bibitem[Hakkila et al. (2008)]{hak08}
Hakkila, J. et al., 2008, \apj, 677, L81
\bibitem[He et al. (2011)]{he11}
He, H.-N., Wu, X.-F., Toma, K., Wang, X.-Y., \& M\'esz\'aros, P. 2011, \apj, 733, 22
\bibitem[Inoue et al.(2011)]{ino11}
Inoue, T., Asano, K., \& Ioka, K. 2011, \apj, 734, 77
\bibitem[Ioka (2010)]{iok10}
Ioka, K. 2010, Prog. Theor. Phys., 124, 667
\bibitem[Ioka at al. (2007)]{iok07}
Ioka, K., Murase, K., Toma, K., Nagataki, S., \& Nakamura, T. 2007, \apj, 670, L77
\bibitem[Ioka \& Nakamura (2001)]{iok01}
Ioka, K., \& Nakamura, T. 2001, \apj, 554, L163
\bibitem[Ioka at al. (2011)]{iok11}
Ioka, K., Ohira, Y., Kawanaka, N., \& Mizuta, A. 2011, arXiv:1103.5746
\bibitem[Kneiske et al. (2004)]{kne04}Kneiske, T.M. et al. 2004, \aap,
413, 807
\bibitem[Kobayashi et al. (1997)]{kob07}
Kobayashi, S., Piran, T., \& Sari, R. 1997, \apj, 490, 92
\bibitem[Kocevski \& Liang (2003)]{koc03}
Kocevski, D., \& Liang, E. 2003, \apj, 594, 385
\bibitem[Kumar \& Barniol Duran (2010)]{kum10}
Kumar, P., \& Barniol Duran, R. 2010, \mnras, 409, 226
\bibitem[Li (2010)]{li10}
Li, Z., 2010, ApJ, 709, 525
\bibitem[Lu et al.(2006)]{lu06}
Lu, R.-J., Qin, Y.-P., Zhang, Z.-B., \& Yi, T.-F. 2006, \mnras, 367, 275
\bibitem[Maxham et al.(2011)]{Max11} Maxham, A., Zhang, B.~B., \&
Zhang, B. 2011, \mnras, 415, 77
\bibitem[M\'esz\'aros (2006)]{mes06}
M\'esz\'aros, P. 2006, Rep. Prog. Phys., 69, 2259
\bibitem[M\'esz\'aros \& Rees (1994)]{mes94}
M\'esz\'aros, P., \& Rees, M. J. 1994, \mnras, 269, L41
\bibitem[M\'esz\'aros \& Rees (2000)]{mes00}
M\'esz\'aros, P., \& Rees, M. J. 2000, \apj, 530, 292
\bibitem[M\'esz\'aros \& Rees (2011)]{mes11}
M\'esz\'aros, P., \& Rees, M. J. 2011, \apjl, 733, L40
\bibitem[Mizuno et al. (2011)]{miz10}
Mizuno, Y., at al. 2011, \apj, 726, 62
\bibitem[Murase et al.(2007)]{mur07}
Murase, K., Asano, K., \& Nagataki, S. 2007, \apj, 671, 1886
\bibitem[Murase \& Nagataki (2006)]{mur06}
Murase, K., \& Nagataki, S. 2006, \prd, 73, 3002
\bibitem[Murase et al.(2011)]{mur11}
Murase, K., Toma, K., Yamazaki, R., \& M\'esz\'aros, P. 2011, \apj, 732, 77
\bibitem[Nakar et al. (2009)]{nak09}
Nakar, E., Ando, S., \& Sari, R. 2009, \apj, 703, 675
\bibitem[Norris et al.(1996)]{nor96}
Norris, J. P. et al., 1996, \apj, 459, 393
\bibitem[Pacz\'ynski (1986)]{pac86}
Pacz\'ynski, B. 1986, \apj, 308, L43
\bibitem[Panaitescu \& M\'esz\'aros (2000)]{pan00}
Panaitescu, A., \& M\'esz\'aros, P. 2000, \apj, 544, L17
\bibitem[Pe'er (2008)]{pee08}
Pe'er, A. 2008, \apj, 682, 463
\bibitem[Pe'er \& Waxman (2005)]{pee05}
Pe'er, A., \& Waxman, E. 2005, \apj, 628, 857
\bibitem[Pe'er \& Zhang (2006)]{pee06}
Pe'er, A., \& Zhang, B. 2006, \apj, 653, 454
\bibitem[Pilla \& Loeb (1998)]{pil98}
Pilla, R. P., \& Loeb, A. 1998, \apj, 494, L167
\bibitem[Piran (2005)]{pir05}
Piran, T. 2005, Rev. Mod. Phys., 76, 1143
\bibitem[Preece et al. (2000)]{pre00}
Preece, R. D., Briggs, M. S., Mallozzi, R. S., Pendleton, G. N., 
Paciesas, W. S., \& Band, D. L. 2000, \apjs, 126, 19
\bibitem[Racusin et al. (2008)]{rac08}
Racusin, J. L. et al., 2008, \nat, 455, 183
\bibitem[Rando et al. (2009)]{ran09}
Rando, R. et al., 2009, arXiv:0907.0626
\bibitem[Rybicki \& Lightman (1979)]{ryb79}Rybicki, G. B., \& Lightman, A. P. 1979,
Radiative Processes in Astrophysics (New York: Wiley-Interscience)
\bibitem[Ryde et al. (2010)]{ryd10}
Ryde, F., et al. 2010, \apj, 709, L172
\bibitem[Shen et al.(2005)]{she05}
Shen, R.-F., Song, L.-M., \& Li, Z. 2005, \mnras, 362, 59
\bibitem[Toma et al. (2009)]{tom09}
Toma, K., Wu, X.-F. \& M\'esz\'aros, P. 2009, \apj, 707, 1404
\bibitem[Toma et al. (2010)]{tom10}
Toma, K., Wu, X.-F. \& M\'esz\'aros, P. 2010, \mnras, 415, 1663
\bibitem[Vietri (1995)]{vie95}
Vietri, M. 1995, \apj, 453, 883
\bibitem[Vurm et al. (2011)]{vur11}
Vurm, I., Beloborodov, A. M., \& Poutanen, J. 2011, accepted for ApJ, arXiv:1104.0394
\bibitem[Vurm \& Poutanen (2009)]{vur09}
Vurm, I., \& Poutanen, J. 2009, \apj, 698, 293
\bibitem[Wang et al. (2009)]{wan09}
Wang, X.-Y., Li, Z., Dai, Z.-G., \& M\'esz\'aros, P. 2009, \apj, 698, 98
\bibitem[Wang \& M\'esz\'aros (2006)]{wan06}
Wang, X.-Y., \& M\'esz\'aros, P. 2006, \apj, 643, L95
\bibitem[Waxman (1995)]{wax95}
Waxman, E. 1995, \prl, 75, 386
\bibitem[Waxman and Bahcall (1997)]{wax97}
Waxman, E., \& Bahcall, J. 1997, Phys. Rev. Lett., 78, 2292
\bibitem[Wu \& Fenimore (2000)]{wu00}
Wu, B., \& Fenimore, E. E. 2000, \apj, 535, L29
\bibitem[Yonetoku et al. (2004)]{yon04}
Yonetoku, D., Murakami, T., Nakamura, T., Yamazaki, R., Inoue, A. K., \&
Ioka, K. 2004, \apj, 609, 935
\bibitem[Zhang et al.(2011)]{zha11}
Zhang, B.-B., et al. 2011, \apj, 730, 141
\bibitem[Zhang et al.(2007)]{zha07}
Zhang, F.-W., Qin, Y.-P., \& Zhang, B.-B. 2007, \pasj, 59, 857
\bibitem[Zhang et al.(2009)]{zha09}
Zhang, W., MacFadyen, A., \& Wang, P. 2009, \apj, 692, 240
\bibitem[Zhao et al.(2011)]{zhao11}
Zhao, X.-H., Li, Z., \& Bai, J.-M. 2011, \apj, 726, 89
\bibitem[Zou et al.(2011)]{zou11}
Zou, Y.-C., Fan, Y.-Z., \& Piran, T. 2011, \apj, 726, 2






\end{thebibliography}
\end{document}